\documentclass[11pt]{article}
\pdfoutput=1
\usepackage{jheppub}
\usepackage{bm,bbm}
\usepackage{booktabs,amsmath}
\usepackage{mathtools}
\usepackage{graphics}
\usepackage{braket}
\usepackage{tikz}
\usepackage{ascmac}
\usepackage{enumitem}
\usepackage{comment}
\usepackage{stackrel}
\usepackage{accents}
\usepackage{cases}
\usepackage{youngtab} 
\usepackage{extarrows}
\usepackage{cancel}
\usetikzlibrary{shapes}
\usepackage{framed}
\usepackage{colortbl}
\definecolor{shadecolor}{rgb}{0.90,0.90,0.90}
\definecolor{color1}{rgb}{0.94,0.86,0.73}
\definecolor{color2}{rgb}{0.94,0.23,0.38}
\definecolor{color3}{rgb}{0.0,0.43,0.72}
\definecolor{color3b}{rgb}{0.0,0.2,0.9}
\definecolor{color1a}{rgb}{0.55,0.62,0.38}
\definecolor{color2a}{rgb}{0.91,0.60,0.61}
\definecolor{color3a}{rgb}{0.52,0.65,0.73}
\definecolor{purple2}{rgb}{0.8,0.0,0.8}
\usetikzlibrary{patterns}
\usetikzlibrary{decorations.markings}
\usetikzlibrary{decorations.pathmorphing}
\tikzset{snake it/.style={decorate, decoration=snake}}

\makeatletter
\newcommand{\mylabel}[2]{#2\def\@currentlabel{#2}\label{#1}}
\makeatother

\usetikzlibrary{quotes,angles}

\usetikzlibrary{shadings}

\tikzset{test/.style n args={3}{
    postaction={
    decorate,
    decoration={
    markings,
    mark=between positions 0 and \pgfdecoratedpathlength step 0.5pt with {
    \pgfmathsetmacro\myval{multiply(
        divide(
        \pgfkeysvalueof{/pgf/decoration/mark info/distance from start}, \pgfdecoratedpathlength
        ),
        100
    )};
    \pgfsetfillcolor{#3!\myval!#2};
    \pgfpathcircle{\pgfpointorigin}{#1};
    \pgfusepath{fill};}
}}}}

\definecolor{c1}{rgb}{0.0,0.2,0.9}
\definecolor{c2}{rgb}{0.45,0.0,0.45}
\definecolor{c3}{rgb}{0.9,0.2,0.0}

\usetikzlibrary{arrows}
\usetikzlibrary {arrows.meta}

\tikzset{middlearrow/.style={
        decoration={markings,
            mark= at position 0.57 with {\arrow{#1}} ,
        },
        postaction={decorate}
    }
}

\usepackage{blkarray}

\usepackage{nicematrix}
\NiceMatrixOptions%
{code-for-first-row = \scriptstyle,
code-for-first-col = \scriptstyle }

\usetikzlibrary{decorations.markings}
\def\MarkLt{4pt}
\def\MarkSep{2pt}

\tikzset{
  TwoMarks/.style={
    postaction={decorate,
      decoration={
        markings,
        mark=at position #1 with
          {
              \begin{scope}[xslant=0.2]
              \draw[line width=\MarkSep,white,-] (0pt,-\MarkLt) -- (0pt,\MarkLt) ;
              \draw[-] (-0.5*\MarkSep,-\MarkLt) -- (-0.5*\MarkSep,\MarkLt) ;
              \draw[-] (0.5*\MarkSep,-\MarkLt) -- (0.5*\MarkSep,\MarkLt) ;
              \end{scope}
          }
       }
    }
  },
  TwoMarks/.default={0.5},
  OneMark/.style={
    postaction={decorate,
      decoration={
        markings,
        mark=at position #1 with
          {
              \draw[-] (0,-\MarkLt) -- (0,\MarkLt) ;
          }
       }
    }
  },
  OneMark/.default={0.5}
}

\edef\restoreparindent{\parindent=\the\parindent\relax}
\usepackage{parskip}
\restoreparindent
\usepackage{ascmac}
\usepackage{mathrsfs}
\usepackage{enumitem}
\newlist{steps}{enumerate}{1}
\setlist[steps, 1]{label = Step \arabic*:}

\renewcommand*\arraystretch{1.2}

\usepackage{tikz}
\usetikzlibrary{intersections, calc,positioning,decorations.pathreplacing,decorations.pathmorphing}
\tikzset{>=latex}

\def\i{{\rm i}}

\def\CC{{\cal C}}
\def\CD{{\cal D}}

\def\CG{{\cal G}}

\def\CR{{\cal R}}
\def\CM{{\cal M}}
\def\CN{{\cal N}}

\def\CS{{\cal S}}
\def\CT{{\cal T}}

\def\CU{{\cal U}}

\def\CZ{{\cal Z}}

\def\BZ{\mathbb{Z}}
\def\BQ{\mathbb{Q}}

\def\b0{\bm{0}_\perp}

\def\U{\text{U}}

\def\SlS{\mathsf{S}}

\usepackage{centernot}
\usepackage{mathtools}
\usepackage{stmaryrd}

\makeatletter
\newcommand{\xMapsto}[2][]{\ext@arrow 0599{\Mapstofill@}{#1}{#2}}
\def\Mapstofill@{\arrowfill@{\Mapstochar\Relbar}\Relbar\Rightarrow}
\makeatother

\usepackage{multirow}

\usepackage{tikz-3dplot}

\DeclareFontFamily{U}{mathx}{\hyphenchar\font45}
\DeclareFontShape{U}{mathx}{m}{n}{
      <5> <6> <7> <8> <9> <10>
      <10.95> <12> <14.4> <17.28> <20.74> <24.88>
      mathx10
      }{}
\DeclareSymbolFont{mathx}{U}{mathx}{m}{n}
\DeclareFontSubstitution{U}{mathx}{m}{n}
\DeclareMathAccent{\widecheck}{0}{mathx}{"71}

\newcommand{\Left}{\text{L}}
\newcommand{\Right}{\text{R}}
\newcommand{\Duality}{{\cal D}}

\newcommand{\BarhatDuality}[1]{\overline{\widehat{{{\cal D}}}}_{#1}}

\newcommand{\hatDuality}[1]{\widehat{\Duality}_{#1}}

\newcommand{\etaSp}{\eta_{\, \SlS\vec{p}}}
\newcommand{\etap}{\eta_{\vec{p}}}
\newcommand{\etaSpb}{(\eta_{\, \SlS\vec{p}})}

\title{Non-invertible duality defect and non-commutative fusion algebra}

\author{Yuta Nagoya and Soichiro Shimamori}
\emailAdd{y\_nagoya@het.phys.sci.osaka-u.ac.jp}
\emailAdd{s\_shimamori@het.phys.sci.osaka-u.ac.jp}

\affiliation{
Department of Physics, Osaka University,\\
Machikaneyama-Cho 1-1, Toyonaka 560-0043, Japan
}

\preprint{OU-HET-1202}

\abstract{
We study non-invertible duality symmetries by gauging a diagonal subgroup of a non-anomalous U(1)$\times$U(1) global symmetry. In particular, we employ the half-space gauging to $c=2$ bosonic torus conformal field theory (CFT) in two dimensions and pure U(1)$\times$U(1) gauge theory in four dimensions. In $c=2$ bosonic torus CFT, we show that the non-invertible symmetry obtained from the diagonal gauging becomes emergent on an \emph{irrational} CFT point. We also calculate the fusion rules concerning the duality defect. We find out that the fusion algebra is \emph{non-commutative}. We also obtain a similar result in pure U(1)$\times$U(1) gauge theory in four dimensions.
}

\begin{document}
\maketitle

\section{Introduction and summary} 
\paragraph{Background:}Global symmetry has always been a pivotal concept in the analysis of quantum field theories (QFTs). One of the most prominent and successful applications of global symmetries is the 't Hooft anomaly matching \cite{tHooft:1979rat}, which aids in our comprehension of the strongly coupled systems. Toward giving further insights into the non-perturbative dynamics of QFTs, the notion of global symmetry has been generalized in \cite{Gaiotto:2014kfa}. There, it has been revealed that a global symmetry is associated with the existence of the {\it topological} defect, and the symmetry transformation can be realized as a boundary condition on the topological defect. 
Although various types of generalized global symmetries have been concerned so far, the non-invertible symmetry has gained significant attention above all\footnote{Historically, non-invertible symmetries have been considered in the context of rational conformal field theories \cite{Verlinde:1988sn,Moore:1988qv,Moore:1989yh,Oshikawa:1996dj,Petkova:2000ip,Fuchs:2002cm,Fuchs:2003id,Fuchs:2004dz,Frohlich:2004ef,Fuchs:2004xi,Fjelstad:2005ua,Frohlich:2006ch,Feiguin:2006ydp,Fuchs:2007tx,Fredenhagen:2009tn,Frohlich:2009gb,Davydov:2010rm,Carqueville:2012dk,Bachas:2013ora,Brunner:2013ota,Brunner:2013xna,Brunner:2014lua,Ho:2014vla,Hauru:2015abi,Aasen:2016dop,Aasen:2020jwb}.}. 
Unlike the ordinary symmetries, the non-invertible symmetry has no inverse operation, hence the resulting fusion algebra forms the fusion category rather than a group \cite{Cui:2016bmd,Douglas:2018,Johnson-Freyd:2020,Kong:2020wmn,Bhardwaj:2022yxj}. In recent years, numerous non-invertible symmetries have been discovered, offering new predictions into the dynamics of QFTs, e.g., constraints on renormalization group flows and realistic QFTs, across diverse dimensions \cite{Bhardwaj:2017xup, Chang:2018iay, Thorngren:2019iar, Thorngren:2021yso, Komargodski:2020mxz,Huang:2021nvb, vanhove2022critical,Heidenreich:2021xpr, Nguyen:2021yld, Nguyen:2021naa, Koide:2021zxj, Choi:2021kmx, Kaidi:2021xfk, Roumpedakis:2022aik, Hayashi:2022fkw, Kaidi:2022uux, Choi:2022jqy, Cordova:2022ieu, Choi:2022zal, Niro:2022ctq, Choi:2022rfe, Choi:2022fgx,Lin:2022dhv, Apruzzi:2022rei,Heckman:2022muc,Heckman:2022xgu,Bhardwaj:2022lsg,Bartsch:2022mpm,Bartsch:2022ytj, Yokokura:2022alv, Apte:2022xtu, Kaidi:2022cpf, Antinucci:2022vyk, Inamura:2022lun, Cordova:2022fhg, Chen:2022cyw, Bashmakov:2022uek, Bhardwaj:2022maz, Bhardwaj:2022kot, Zhang:2023wlu, Kaidi:2023maf,Lin:2023uvm,Seiberg:2023cdc,Cao:2023doz,Bhardwaj:2023ayw,Bartsch:2023wvv, Choi:2023xjw, Koide:2023rqd, Inamura:2023qzl, Apruzzi:2023uma,vanBeest:2023dbu,Haghighat:2023sax,Chen:2023czk, Cordova:2023ent, Damia:2023ses,Lawrie:2023tdz, Antinucci:2023ezl,Pace:2023kyi, Cordova:2023bja, Valentin:2023,Choi:2023pdp}. (See \cite{Schafer-Nameki:2023jdn, Shao:2023gho} for the comprehensive review on the non-invertible symmetry.)
\begin{figure}
    \centering
 \begin{tikzpicture}[scale=0.7]
 \fill[blue!20] (5, 0)--(10, 0)--(10, 5)--(5, 5)--cycle;
\draw[black!80, very thick] (0, 0)--(10, 0)--(10, 5)--(0, 5)--(0, 0);
\draw[color2, ultra thick] (5, 0)--(5, 5); 
\node[black] at (2.5, 2.5) {$\CT$}; 
\node[black] at (7.5, 2.5) {$\CT/H\cong \CT$}; 
  \end{tikzpicture}
    \caption{Pictorical representation of the half-space gauging.}
    \label{fig: half-space gauging c=1}
\end{figure}
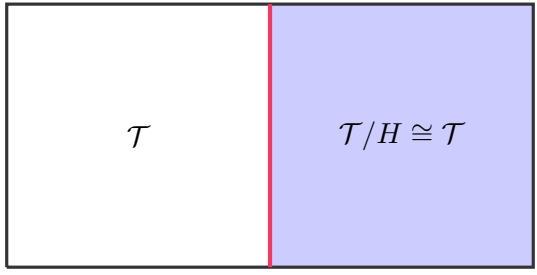

\paragraph{Half-space gauging:}In these developments, the half-space gauging plays a crucial role in systematically constructing the non-invertible duality defects \cite{Choi:2021kmx, Kaidi:2021xfk}. To consider half-space gauging, we first split the spacetime manifold $X$ into the left and right regions separated by the co-dimension one interface as depicted in Figure \ref{fig: half-space gauging c=1}. We perform gauging a non-anomalous discrete global symmetry $H$ of a theory $\CT$ only in half of the spacetime and impose the Dirichlet boundary condition on the $H$ gauge field at the interface. Then, for some special cases, the theory becomes invariant under gauging $H$: $\CT/H\cong \CT$, and the interface becomes (topological) non-invertible defect $\CN$. Here, let us briefly summarize the non-invertible symmetries constructed from the half-space gauging in $c=1$ compact boson model \cite[section 4.1]{Choi:2021kmx};
\begin{align}
    \frac{R^2}{4\pi}\int_{X_{2}}d\phi\,  \wedge\,  \star d\phi\ , 
\end{align}
where $X_{2}$ is a two-dimensional orientable manifold, and $\phi$ is the compact boson with the periodicity
$2\pi$. In this case, we gauge the discrete shift symmetry $\BZ_{N}\subset$U$(1)^{\text{shift}}$ only in the right region. By using T-duality, we can achieve $\CT/\BZ_{N}\cong \CT$ only if we tune the compact radius such that $R=\sqrt{N}$, which describes the rational conformal field theory (RCFT). Notably, the non-invertible duality defect $\CN$ can be expressed by the following action;
\begin{align}
    \CN\ :\ \i\frac{ N}{2\pi}\int_{x=0} \phi_{\text{L}}d\phi_{\text{R}}\ , 
\end{align}
where $\phi_{\text{L}}$ and $\phi_{\text{R}}$ are the compact boson fields that live in the left and right regions, respectively. The fusion algebra concerning the non-invertible duality defect $\CN$ and the $\BZ_{N}$ shift symmetry generator $\eta$ are given by the following Tambara-Yamagami category \cite{TAMBARA1998692}; 
\begin{align}\label{eq: Tambara-Yamagami c=1}
    \begin{aligned}
         \CN\times \CN&=\CC \ , \\
        \eta\times \CN&=\CN\times \eta=\CN\ , \\
         \eta^N &=1\ , 
     \end{aligned}
 \end{align}
 where $\CC=1+\eta+\eta^2 +\cdots \eta^{N-1}$ is the projection operator of the $\BZ_{N}$ shift symmetry up to normalization.
In summary, in $c=1$ compact boson CFT, the non-invertible symmetry obtained by the half-space gauging becomes emergent at $R=\sqrt{N}$, namely RCFT point, and the fusion algebra is given by the Tambara-Yamagami category \eqref{eq: Tambara-Yamagami c=1}.

\paragraph{Motivations:}The most natural and simplest generalization of the above $c=1$ compact boson CFT is the $c=2$ bosonic torus CFT;\footnote{For simplicity, we do not include the topological term in the action \eqref{eq: c=2 cft action} in this paper.}
\begin{align}\label{eq: c=2 cft action}
    \frac{1}{4\pi}\int_{X_{2}}\,G_{IJ}\,  d\phi^{I}\wedge\star \, d\phi^{J}\ \qquad , \qquad I, J=1\, , 2\ ,  
\end{align}
where $\phi^{I}$ is the compact boson with periodicity $2\pi$. Then, inspired by the above example of the $c=1$ compact boson theory, the following two questions naturally arise;
\begin{itemize}
    \item Where do the non-invertible symmetries obtained from the half-space gauging become emergent on the conformal manifold? In particular, are these non-invertible symmetries found at rational or irrational CFT points?
    \item What is the fusion algebra associated with the non-invertible symmetry defect?
\end{itemize}
The main aim of this paper is to address these questions. As we will see later, the landscape of non-invertible symmetries from the half-space gauging in the $c=2$ bosonic torus CFT is richer than the $c=1$ case. We also apply the half-space gauging to the pure U(1)$\times$U(1) gauge theory in four dimensions;
\begin{align}
        \frac{1}{4\pi}\int_{M_{4}}\,\CG_{IJ}\,  dA^{I}\wedge\star \, dA^{J}\ \qquad , \qquad I, J=1\, , 2\ ,
\end{align}
and investigate the non-invertible structures of this theory. In the remainder of the Introduction, we present a concise summary of our work.

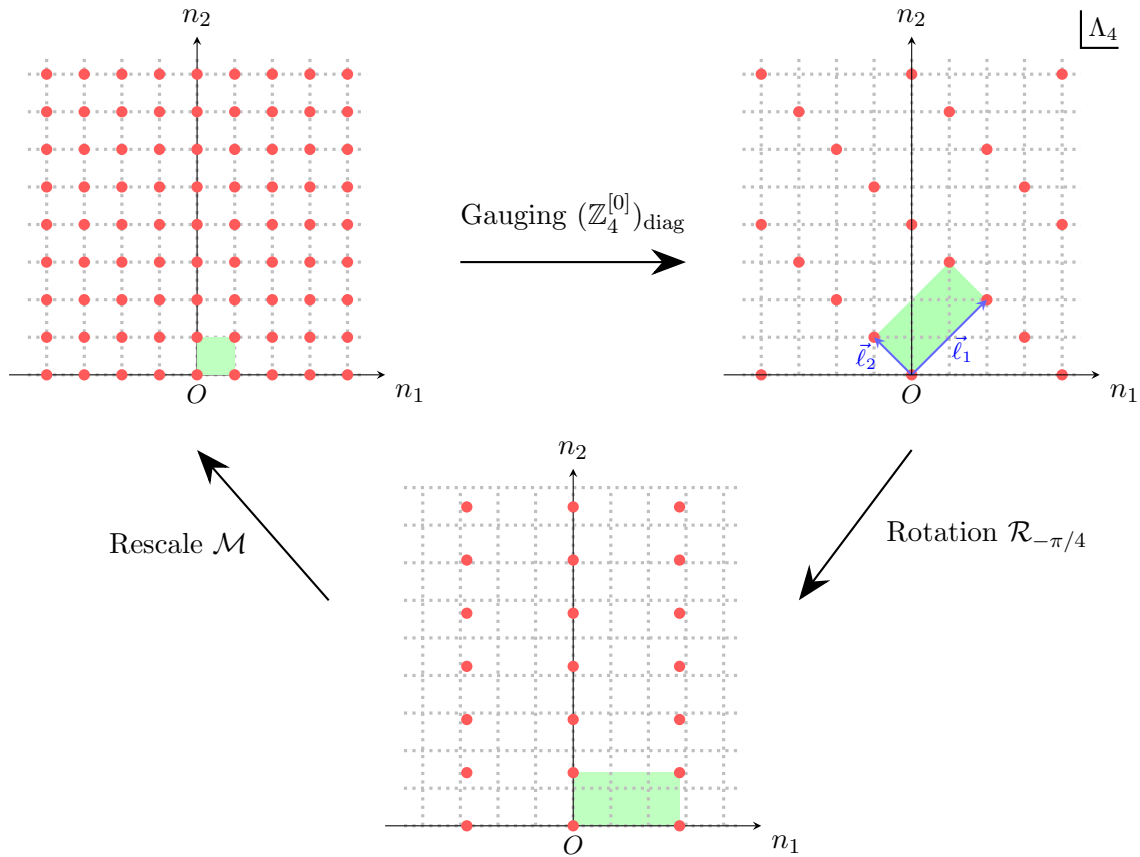
\begin{figure}[t]
     \centering
  \begin{tikzpicture}[scale=0.5]
    grid size
   \def\N{8}
\fill[green!30] (23, 0)--(25,2)--(24, 3)--(22, 1)--(23, 0);
   \draw[step=1, lightgray, dotted, very thick] (-0.5, 0) grid (\N+0.5, \N+0.5);

  \draw[arrows=-{stealth[]}] (-1,0) -- (\N+1,0) node[below right] {$n_1$};
  \draw[arrows=-{stealth[]}] ({\N/2},0) node[below]{{\small $O$}} -- ({\N/2},\N+1) node[above] {$n_2$};
  \fill[green!25] (4, 0)--(4,1)--(5, 1)--(5, 0)--(4, 0);
  \foreach \x in {0,...,\N} {
    \foreach \y in {0,...,\N} {
      \fill[red!65] (\x, \y) circle (0.15);
    }
  }
  \draw [arrows = {-Stealth[scale=2]}, thick] (11,3) -- (17,3); \node[above] at (14, 3.5) {Gauging $(\BZ^{[0]}_{4})_{\text{diag}}$};

  \draw[step=1, lightgray, dotted, very thick] (19-0.5, 0) grid ({\N+19+0.5}, \N+0.5);
  
\def\gridpoints{{0,0}, {0,4}, {0,8}, {1,3}, {1,7}, {2,2}, {2,6}, {3,1}, {3,5}, {4, 0}, {4,4}, {4,8}, {5,3}, {5,7}, {6,2}, {6,6},
                {7,1}, {7,5}, {8, 0}, {8,4}, {8,8}}

  \fill[red!65] (19, 0) circle (0.15);
  \fill[red!65] (19, 4) circle (0.15);
  \fill[red!65] (19, 8) circle (0.15);
  \fill[red!65] (20, 3) circle (0.15);
  \fill[red!65] (20, 7) circle (0.15);
  \fill[red!65] (21, 2) circle (0.15);
  \fill[red!65] (21, 6) circle (0.15);
  \fill[red!65] (22, 1) circle (0.15);
  \fill[red!65] (22, 5) circle (0.15);
  \fill[red!65] (23, 0) circle (0.15);
  \fill[red!65] (23, 4) circle (0.15);
  \fill[red!65] (23, 8) circle (0.15);
  \fill[red!65] (24, 3) circle (0.15);
  \fill[red!65] (24, 7) circle (0.15);
  \fill[red!65] (25, 2) circle (0.15);
  \fill[red!65] (25, 6) circle (0.15);
  \fill[red!65] (26, 1) circle (0.15);
  \fill[red!65] (26, 5) circle (0.15);
  \fill[red!65] (27, 0) circle (0.15);
  \fill[red!65] (27, 4) circle (0.15);
  \fill[red!65] (27, 8) circle (0.15);
  \draw[arrows=-{stealth[]}, thick, blue!60] (23,0) -- (25,2);
  \draw[arrows=-{stealth[]}, thick, blue!60] (23,0) -- (22,1);
  \draw[arrows=-{stealth[]}] ({-1+19},0) -- ({\N+1+19},0) node[below right] {$n_1$};
  \draw[arrows=-{stealth[]}] ({\N/2+19},0) node[below]{{\small $O$}} -- ({\N/2+19},\N+1) node[above] {$n_2$};
\node[below right] at (23.8,1.3) {\textcolor{blue}{{\footnotesize $\vec{\ell}_1$}}};
\node at (21.8,0.5) {\textcolor{blue}{{\footnotesize $\vec{\ell}_2$}}};
\draw[thick] (27.5, 9.7)--(27.5, 8.7)--(28.5, 8.7);
\node at (28.1, 9.2) {{\small $\Lambda_{4}$}};

\draw [arrows = {-Stealth[scale=2]}, thick] (23, -2) -- (20,-6); \node[above] at (25, -5) {Rotation $\CR_{-\pi/4}$};

 \fill[green!25] (14, -12)--({2*sqrt(2)+14},-12)--({2*sqrt(2)+14},{sqrt(2)-12})--(14,{sqrt(2)-12})--(14, -12);
  \draw[step=1, lightgray, dotted, very thick] (9.5, -12) grid (18.5, -3);
  \draw[arrows=-{stealth[]}] ({-1+10},-12) -- ({\N+1+10},-12) node[below right] {$n_1$};
  \draw[arrows=-{stealth[]}] ({\N/2+10},-12) node[below]{{\small $O$}} -- ({\N/2+10},\N+1-11.5) node[above] {$n_2$};

\foreach \x in {-1,...,1} {
    \foreach \y in {0,...,6} {
      \fill[red!65] ({2*sqrt(2)*\x+14}, {sqrt(2)*\y-12}) circle (0.15);
    }
  }

\draw [arrows = {-Stealth[scale=2]}, thick] (7.5, -6) -- (4,-2); \node[above] at (3.5, -5) {Rescale $\CM$};
  \end{tikzpicture}
    \caption{The transition of the charged lattice corresponding to the gauging $(\BZ^{[0]}_{4})_{\text{diag}}$. The horizontal/vertical axis denotes the U$(1)^{\text{shift}}_{1}$/ U$(1)^{\text{shift}}_{2}$ charge of the vertex operator $e^{\i\vec{n}\cdot\vec{\phi}}$, respectively. The red points in each diagram mean the properly quantized charges and the green realm shows the unit cell. In order to bring the charge lattice after gauging $(\BZ^{[0]}_{4})_{\text{diag}}$ back to the original one $\BZ\times\BZ$, we need to perform rotation and rescaling as depicted in the figure.}
    \label{fig: transition of charged lattice intro}
\end{figure}

\paragraph{Summary:}
 The $c=2$ bosonic torus CFT has the zero-form shift-symmetry $\text{U}(1)_{1}^{\text{shift}}\times\text{U}(1)_{2}^{\text{shift}}$. Its charged operator is the vertex operator $e^{\i \vec{n}\cdot\vec{\phi}}$ which is characterized by two integers: $\vec{n}\in \BZ\times \BZ$. Under U$(1)_{1}^{\text{shift}}\times$U$(1)_{2}^{\text{shift}}$, the vertex operator is transformed in the following way;
  \begin{align}
     \text{U(1)}_{1}^{\text{shift}}\times \text{U(1)}_{2}^{\text{shift}}\ : \ e^{\i \vec{n}\cdot\vec{\phi}}\mapsto e^{\i \vec{\theta} \cdot \vec{n}}\, e^{\i \vec{n}\cdot\vec{\phi}} \qquad  , \qquad \theta^1 , \theta^2 \in [0, 2\pi)\ . 
 \end{align}
As a discrete subgroup of the shift symmetry to be gauged, we choose the diagonal subgroup $(\BZ^{[0]}_{2N'})_{\text{diag}}$, whose generator is specified by $(\,e^{\i \frac{ \pi}{N'}}, e^{\i  \frac{\pi}{N'}} \, )$. As a result of the gauging $(\BZ^{[0]}_{2N'})_{\text{diag}}$, the charge lattice of the original theory $\BZ\times \BZ$ is reduced to its sublattice $\Lambda_{2N'}$ defined by;
\begin{align}\label{eq: charge lattice gauging condition}
\begin{aligned}
        \Lambda_{2N'}&\equiv \{\vec{n}\in \BZ\times \BZ\mid  n_1 + n_2 =0\mod 2N' \} \\
        &=\text{Span}(\vec{\ell}_{1}, \vec{\ell}_{2})\ , 
\end{aligned} 
\end{align} 
where the charge lattice $\Lambda_{2N'}$ is spanned by the two orthogonal vectors $\vec{\ell}_{1}$ and $\vec{\ell}_{2}$ (see the upper right lattice in Figure \ref{fig: transition of charged lattice intro}.);
\begin{align}
    \vec{\ell}_{1}
    =\begin{pmatrix}
        N' \\
        N' 
    \end{pmatrix}\qquad , \qquad 
    \vec{\ell}_{2}
    =\begin{pmatrix}
        -1 \\
        1 
    \end{pmatrix}\ .
\end{align}
For later convenience, we put these two basis vectors $\vec{\ell}_1$ and $\vec{\ell}_2$ into the matrix $K$ defined by;\footnote{Throughout this paper, we choose the charge matrix $K$ such that $\det K=+2N'$.}
\begin{align}\label{eq: charge matrix intro}
    K\equiv(\vec{\ell}_1 , \vec{\ell}_2) =\begin{pmatrix}
        N'& -1  \\
        N' & 1
    \end{pmatrix}\ ,
\end{align}
which clearly carries an information of the charge lattice $\Lambda_{2N'}$. In order for the theory to be invariant under the diagonal gauging $(\BZ^{[0]}_{2N'})_{\text{diag}}$, we must perform the rotation and rescaling on the charge lattice $\Lambda_{2N'}$ and bring it back to the original one $\BZ\times \BZ$. The charge lattice transition under these operations is depicted in Figure \ref{fig: transition of charged lattice intro}.
Note that the rotation is a peculiar operation to the $c=2$ bosonic torus CFT. 
Moreover, by using the T-duality, we can perfectly restore the original theory $\CT/(\BZ^{[0]}_{2N'})_{\text{diag}}\cong \CT$, if the kinetic matrix $G_{IJ}$ is tuned to satisfy the following self-duality condition;
\begin{align}\label{eq: self-duality condition intro}
    G=K^{\text{T}} G^{-1} K\ .
\end{align}
The solution $G_{IJ}^{*}$ to the self-duality condition \eqref{eq: self-duality condition intro} is given by
\begin{align}
    G^{*} =\sqrt{-\frac{N}{D}}
    \begin{pmatrix}
        2K_{11} & K_{12}+K_{21} \\
   K_{12}+K_{21} & 2K_{22}
    \end{pmatrix}\quad , \quad D\equiv (K_{12}+K_{21})^{2}-4K_{11}K_{22}\ ,
\end{align}
and we can show that the solution $G_{IJ}^*$ corresponds to the complex multiplication (CM) point\footnote{Here, we call the CM point when either the complex structure modulus $\tau$ {\it or} the complexified K\"{a}hler modulus $\rho$ (see \eqref{eq: complex structure modulus} and \eqref{eq: Kahler moduli} for their definitions) belongs to an imaginary quadratic field. If these two moduli are elements of the {\it same} imaginary quadratic field, the CM point is enhanced to the RCFT one.} \cite{Gukov:2002nw}, which is a wider class of the $c=2$ bosonic torus RCFTs. We also prove that the non-invertible symmetry constructed from the half-space gauging associated with the diagonal gauging becomes emergent on the {\it irrational} CFT.

Furthermore, we also show that the non-invertible symmetry defect $\CD$ associated with the gauging $(\BZ^{[0]}_{2N'})_{\text{diag}}$ can be put into the following Lagrangian form;
\begin{align}\label{eq: defect action duality intro}
    \CD\ : \ \frac{\i}{2\pi}\int_{x=0}K_{IJ}\phi_{\text{L}}^{I}\, d\phi_{\text{R}}^{J}\ , 
\end{align}
and derive the fusion algebra. We find that the resulting fusion algebra is \emph{infinitely} generated and \emph{non-commutative}. We also discover the closed fusion subalgebra. To see this, we must put the various global symmetry generators on the duality defect $\CD$, and define the {\it dressed} duality defect $\widehat{\CD}_{s_1, s_2}$ ($s_1 , s_2 =0, 1, \cdots 2N'-1$). (See section \ref{sec:Fusion rules} for the definition.) Thereby, the projection operator should also be replaced by the dressed one $\widehat{\CC}_{s_1, s_2}$. As a result of this dressing, the fusion algebra concerning $\widehat{\CD}_{s_1, s_2}\,$, $\widehat{\CC}_{s_1, s_2}\,$, and the $(\BZ^{[0]}_{2N'})_{\text{diag}}$ shift symmetry generator $\eta_{\vec{p}}$ can be summarized as follows; 
\begin{shaded}
    \noindent \textbf{Non-commutative fusion subalgebra at the irrational CFT point}
    \begin{align}\label{eq: fusion category in Intro}
        \begin{aligned}
        &\hatDuality{s_1,s_2}\times \hatDuality{s_3,s_4}=\widehat{\CC}_{s_2+s_3,s_1+s_4}\ ,\\
      &\hatDuality{s_1,s_2}\times\eta_{\vec{p}}=\eta_{\vec{p}}\times\hatDuality{s_1,s_2}=\hatDuality{s_1,s_2}\ ,\\
      &\widehat{\CD}_{s_1,s_2}\times \widehat{\CC}_{s_3,s_4}=2N'\, \hatDuality{s_1+s_3,s_2+s_4} \ , \\ &\widehat{\CC}_{s_1,s_2}\times \widehat{\CD}_{s_3,s_4}=2N'\, \hatDuality{s_2+s_3,s_1+s_4}\ , \\
            &\widehat{\CC}_{s_1,s_2}\times\widehat{\CC}_{s_3,s_4}=2N'\, \widehat{\CC}_{s_1+s_3,s_2+s_4}\ ,\\
        &\eta_{\vec{p}}^{2N'}=1\ .
        \end{aligned}
    \end{align}
    \end{shaded}
We also consider the half-space gauging with respect to the product group $\BZ_{N_1}\times\BZ_{N_2}$, instead of the diagonal one. In this case, the non-invertible symmetry arises on the RCFT point, and the fusion algebra is given by the standard Tambara-Yamagami category \eqref{eq: Tambara-Yamagami c=1} since $c=2$ bosonic torus CFT is reduced to the two sets of the $c=1$ compact boson CFTs. 
Our main results described above are summarized in Table \ref{table: our results}.

Finally, we apply the half-space gauging and explore the non-invertible symmetries in the pure U(1)$\times$U(1) gauge theory in four dimensions.
This theory has the U(1)$^{\text{ele}}_1\times$U(1)$^{\text{ele}}_2$ electric one-form symmetry, whose charged object is the Wilson loop. In a similar manner to the two dimensions, we construct the non-invertible symmetries from gauging the diagonal subgroup $(\BZ_{N}^{[1]})_{\text{diag}}\subset$ U(1)$^{\text{ele}}_1\times$U(1)$_2^{\text{ele}}$. By utilizing the electric-magnetic duality transformation, we find out the special gauge couplings where the non-invertible symmetries appear.
 As in the two dimensions, we construct the duality defect associated with the diagonal gauging and calculate the fusion rules concerning the duality defect. The resulting fusion algebra is again \emph{infinitely} generated and \emph{non-commutative}. We also find out the closed fusion subalgebra which is mostly similar to \eqref{eq: fusion category in Intro}. It remains an open question how we interpret the obtained non-commutative fusion algebra in the framework of the higher category \cite{Bhardwaj:2022yxj}. 

\begin{table}[t]
\renewcommand{\arraystretch}{2.2}
\centering
\begin{tabular}{>{\centering}m{3.8cm}>{\centering}m{3.0cm}>{\centering}m{3.5cm}>{\centering\arraybackslash}m{4cm}}
\toprule
    Gauging group&Charge matrix& Emergent point& 
    Fusion algebra 
    \\
  \midrule 
  Diagonal group $(\BZ^{[0]}_{2N'})_{\text{diag}}$&$K^{\text{T}}\not= K$&Irrational CFT & Non-commutative \eqref{eq: fusion category in Intro}  \\
    \hline
    Product group $\BZ_{N_1}^{[0]}\times \BZ_{N_2}^{[0]}$ & $K^\text{T}=K$ &RCFT& Tambara-Yamagami \eqref{eq: Tambara-Yamagami c=1} \\
    \bottomrule
\end{tabular}
\caption{Summary of the main results in $c=2$ bosonic torus CFT.  
}
\label{table: our results}
\end{table}

The rest of the paper is organized as follows. In section \ref{sec: general theory}, we describe our method to construct non-invertible symmetries from the half-space gauging in arbitrary even dimensions. In particular, we give a detailed explanation of each step from the diagonal gauging to the rotation and the rescaling of the charge lattice, to the duality. In section \ref{sec:Example in two dimensions: $c=2$ bosonic torus CFT}, we discuss the non-invertible symmetries in $c=2$ bosonic torus CFT. In section \ref{subsubsec: self-duality condition}, we derive the self-duality condition \eqref{eq: self-duality condition intro}, and show that the solution corresponds to the CM point. 
We also show that the non-invertible symmetry associated with the diagonal gauging $(\BZ^{[0]}_{2N'})_{\text{diag}}$ becomes emergent at the {\it irrational} CFT point.
In section \ref{subsec: Duality defect}, we construct the duality defect action \eqref{eq: defect action duality intro}, and describe various aspects of the duality defect e.g., the boundary condition on the defect, topological property, and the orientation reversion. In section \ref{sec:Fusion rules}, we elucidate the precise definition of the dressed duality defect $\widehat{\CD}_{s_1 , s_2}$ and discuss the fusion algebra. 
In section \ref{sec: Example in four dimensions}, we consider the pure U(1)$\times$U(1) gauge theory in four dimensions and explain that the non-invertible symmetries from the half-space gauging from the diagonal gauging can be constructed in a very similar manner to the two-dimensions. Furthermore, we show that the resulting fusion algebra is also non-commutative, and end with giving an open question on our fusion algebra. In section \ref{sec: conclusion and discussion}, we briefly summarize this paper and discuss the future directions. In appendix \ref{sec: derivation of fusion algebra}, we derive some selected fusion rules, skipped in the main text. 

\section{Non-invertible symmetry from half-space gauging $(\BZ^{[q]}_{2N'})_{\text{diag}}$}\label{sec: general theory}
\begin{table}[t]
\label{table:2 and 4 dimensions}
\renewcommand{\arraystretch}{2.2}
\centering
\begin{tabular}{>{\centering}m{1.8cm}>{\centering}m{3.2cm}>{\centering}m{3cm}>{\centering\arraybackslash}m{2.7cm}>{\centering\arraybackslash}m{2.7cm}}
\toprule
    & Theory & 
    U$(1)_{1}^{[q]}\times$U$(1)_{2}^{[q]}$ & Charged ops. $V_{\vec{n}}^{[q]}$ & Duality
    \\
  \midrule 
  $d=2$ \\$(q=0)$&{\small $c=2$ bosonic torus CFT (section \ref{sec:Example in two dimensions: $c=2$ bosonic torus CFT})} & {\small U$(1)^{\text{shift}}_{1}\times$U$(1)^{\text{shift}}_{2}$}&$e^{\i \vec{n}\cdot \vec{\phi}}$& T-duality  \\
    \hline
   $d=4$ \\$(q=1)$&{\small U(1)$\times$U(1) gauge theory (section \ref{sec: Example in four dimensions})} & {\small  U$(1)^{\text{ele}}_{1}\times$U$(1)^{\text{ele}}_{2}$}&$\displaystyle e^{\i \vec{n}\cdot \int_{\gamma} \vec{A}}$& {\footnotesize Electric-Magnetic duality}  \\
    \bottomrule
\end{tabular}
\caption{Examples treated in this paper and the corresponding notations with section \ref{sec: general theory}.
Detailed explanations for the two-dimensional and four-dimensional examples are deferred to section \ref{sec:Example in two dimensions: $c=2$ bosonic torus CFT} and \ref{sec: Example in four dimensions}, respectively.
}
\label{table: notations}
\end{table}
In this section, we describe the general method to construct the non-invertible symmetry of the theory $\CT_{g}$ whose collective couplings are symbolically denoted by $g$. We assume that the global symmetry contains a non-anomalous $q=\frac{d-2}{2}$ form symmetry $\text{U}(1)^{[q]}_{1}\times \text{U}(1)^{[q]}_{2}$ whose $q$ dimensional charged object is denoted by $V^{[q]}_{\vec{n}}$. 
The charged operator $V_{\vec{n}}$ transforms as;
\begin{align}\label{eq: U(1) trsf}
    \begin{aligned}
        \U(1)^{[q]}_{1}\times\U(1)^{[q]}_{2} &:\quad  V^{[q]}_{\vec{n}}\mapsto e^{\i \vec{\theta}\cdot \vec{n}}\, V^{[q]}_{\vec{n}}\, ,
    \end{aligned}
\end{align}
where $\vec{\theta}\equiv(\theta^{1}, \theta^{2})$ are rotational angles. Importantly, in order for the $2\pi$ rotation to be trivial, the U$(1)^{[q]}_{1}\times$U$(1)^{[q]}_{2}$ charge $\vec{n}$ must be quantized to be integers;
\begin{align}
    \vec{n}\in \BZ\times \BZ\ . 
\end{align}
We refer to the set of properly quantized charges of $V^{[q]}_{\vec{n}}$ as the {\it charge lattice}. In this sense, the original theory $\CT_{g}$ has the charge lattice $\BZ\times \BZ$. (See Table \ref{table:2 and 4 dimensions} for referring examples treated in this paper.)

Our construction of the non-invertible symmetry in the theory $\CT_{g}$ can be schematically summarized as the following diagram;
\begin{align}\label{eq: strategy for constuction}
\begin{aligned}
        &\CT_{g}&\overset{\text{Gauging }(\BZ^{[q]}_{2N'})_{\text{diag}}}{\xlongrightarrow[]{\hspace{4em}}}\ \CT_{g}/(\BZ^{[q]}_{2N'})_{\text{diag}}\ &\overset{\CR_{-\theta} \text{ and } \CM}{\xlongrightarrow[]{\hspace{4em}}}\ \CT_{g'}'\ \overset{\text{Duality}}{\xlongrightarrow[]{\hspace{4em}}}\ \widehat{\CT'}_{\widehat{g}\,'}\cong\CT_{g}\ .
\end{aligned}
\end{align}
In particular, the transition of the charge lattice at each step is shown below;
\begin{align}\label{eq: transition of charge lattice}
    \BZ\times \BZ \overset{\text{Gauging }(\BZ^{[q]}_{2N'})_{\text{diag}}}{\xlongrightarrow[]{\hspace{4em}}}\ \Lambda_{2N'} \ \overset{\CR_{-\theta} \text{ and } \CM}{\xlongrightarrow[]{\hspace{4em}}}\ \CM\CR_{-\theta} \, \Lambda_{2N'}\cong\BZ\times \BZ \ \overset{\text{Duality}}{\xlongrightarrow[]{\hspace{4em}}}\ \widehat{\BZ\times \BZ}\cong\BZ\times \BZ \ . 
\end{align}
In the rest of this section, we provide a detailed explanation of each step in \eqref{eq: strategy for constuction} and \eqref{eq: transition of charge lattice}. 

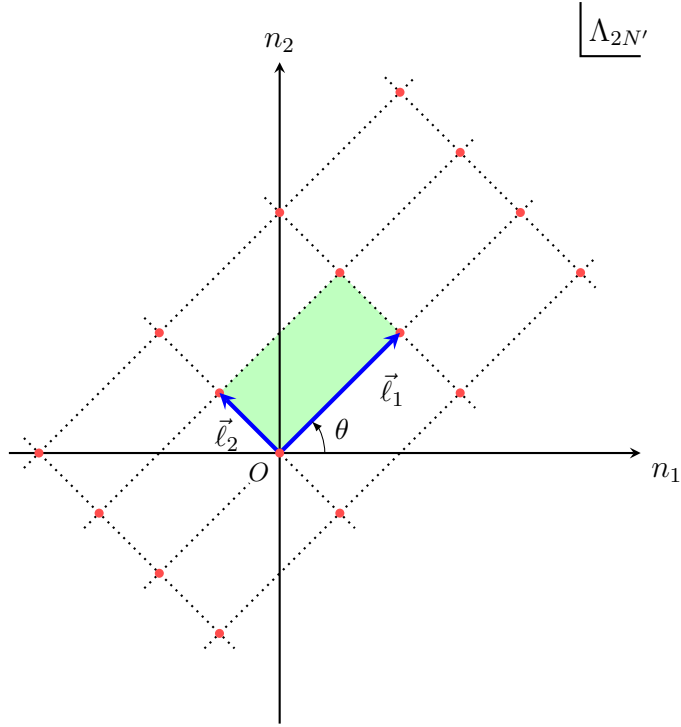
\begin{figure}[t]
    \centering
 \begin{tikzpicture}[scale=0.4]

\fill[green!25] (0, 0)--(4,4)--(2, 6)--(-2, 2)--(0, 0);
\draw[dotted, thick] (-8.5, -0.5)--(4.5, 12.5); 
\draw[dotted, thick] (-6.5, -2.5)--(6.5, 10.5);
\draw[dotted, thick] (-4.5, -4.5)--(-1, -1);
\draw[dotted, thick] (0, 0)--(8.5, 8.5);
\draw[dotted, thick] (-2.5, -6.5)--(10.5, 6.5);
\draw[dotted, thick] (-8.5, 0.5)--(-1.5, -6.5); 
\draw[dotted, thick] (-4.5, 4.5)--(2.5, -2.5);
\draw[dotted, thick] (-0.5, 8.5)--(6.5, 1.5);
\draw[dotted, thick] (3.5, 12.5)--(10.5, 5.5);
\draw[thick] (10, 15)--(10, 13.2)--(12, 13.2);
\node at (11.3, 14) {{\large $\Lambda_{2N'}$}};
   \draw[arrows=-{stealth[]}, thick] (-9,0) -- (12,0) node[below right] {{\large $n_1$}};
  \draw[arrows=-{stealth[]}, thick] (0,-9) -- (0,13) node[above] {{\large $n_2$}};
  \fill[red!70] (4, 4) circle (0.15);
  \fill[red!70] (2, 6) circle (0.15);
  \fill[red!70] (-2, 2) circle (0.15);
  \fill[red!70] (-4, 4) circle (0.15);
  \fill[red!70] (0, 8) circle (0.15);
  \fill[red!70] (8, 8) circle (0.15);
  \fill[red!70] (6, 10) circle (0.15);
  \fill[red!70] (4, 12) circle (0.15);
  \fill[red!70] (10, 6) circle (0.15);
  \fill[red!70] (6, 2) circle (0.15);
  \fill[red!70] (-8, 0) circle (0.15);
   \fill[red!70] (-6, -2) circle (0.15);
   \fill[red!70] (-4, -4) circle (0.15);
   \fill[red!70] (-2, -6) circle (0.15);
   \fill[red!70] (2, -2) circle (0.15);
    \draw[arrows=-{stealth[]}, ultra thick, blue] (0,0) -- (4,4);
  \node[below right] at (3,3) {$\vec{\ell}_1$};
  \draw[arrows=-{stealth[]}, ultra thick, blue] (0,0) -- (-2,2);
  \node[below left] at (-1,1.5) {$\vec{\ell}_2$};
  \draw [->](1.5,0) arc [start angle = 0, end angle = 45, radius = 1.5];
  \node at (1.5, 0.8) [right] {$\theta$};
   \fill[red!70] (0, 0) node[below left, black]{{\small $O$}} circle (0.15);
  \end{tikzpicture}
    \caption{Picture of the charge lattice $\Lambda_{2N'}$. The elements of the charge lattice $\Lambda_{2N'}$ are depicted by red points, with the origin denoted by $O$. The unit cell is shown by the green realm, and its orthogonal vectors $\vec{\ell}_{1}$ and $\vec{\ell}_{2}$ are represented by two blue arrows. Also, for later convenience, we introduce the angle $\theta$ between the first axis and the vector $\vec{\ell}_{1}$.}
    \label{fig: unit cell of L}
\end{figure}
\subsection{Gauging $(\BZ^{[q]}_{2N'})_{\text{diag}}$} 
    The story begins with gauging the diagonal discrete subgroup
    $(\BZ^{[q]}_{2N'})_{\text{diag}}\subset \text{U}(1)^{[q]}_{1}\times \text{U}(1)^{[q]}_{2}$. Here, the diagonal group $(\BZ^{[q]}_{2N'})_{\text{diag}}$ is generated by $(\,e^{\i \frac{\pi}{N'}}, e^{\i \frac{\pi}{N'}} \, )$.
From \eqref{eq: U(1) trsf}, the charged operator $V^{[q]}_{\vec{n}}$ is transformed  under $(\BZ^{[q]}_{2N'})_{\text{diag}}$ as follows;
\begin{align}\label{eq: act of ZNmixed}
    (\BZ^{[q]}_{2N'})_{\text{diag}}:\  V^{[q]}_{\vec{n}} \mapsto e^{\i\frac{\pi}{N'}\vec{p}\cdot \vec{n}}\, V^{[q]}_{\vec{n}}\qquad ,  \qquad \vec{p}=(1,1)^{\text{T}}\ .
\end{align}
Therefore, as a consequence of the diagonal gauging, the charge lattice of the original theory $\BZ\times \BZ$ is projected out to its sublattice $\Lambda_{2N'}$ defined by;
\begin{align}\label{eq: charge lattice gauging condition}
    \Lambda_{2N'}\equiv \{\vec{n}\in \BZ\times \BZ\mid  n_1 + n_2 =0\mod 2N' \}\ , 
\end{align}
which is depicted in Figure \ref{fig: unit cell of L}. The new charge lattice $\Lambda_{2N'}$ is spanned by the two orthogonal vectors $\vec{\ell}_{1}$ and $\vec{\ell}_{2}$;
    \begin{align}
            \vec{\ell}_{1}=\begin{pmatrix}
   N' \\
   N'  
\end{pmatrix}\qquad , \qquad 
    \vec{\ell}_{2}=\begin{pmatrix}
   -1 \\
   1  
\end{pmatrix}\ ,  
    \end{align}
and we define the charge matrix $K$ as follows;
\begin{align}\label{eq: charge matrix}
    K\equiv(\vec{\ell}_1 , \vec{\ell}_2) =\begin{pmatrix}
        N'& -1  \\
        N' & 1
    \end{pmatrix}\ .
\end{align}
\subsection{Rotation and rescaling} 
As a result of the diagonal gauging, the charge lattice $\Lambda_{2N'}$ is clearly different from the original one $\BZ\times \BZ\,$. Therefore, in order to construct a symmetry, the charge lattice $\Lambda_{2N'}$ must be restored to the original one $\BZ\times \BZ$. To achieve this, two operations are needed. Firstly, the charge lattice $\Lambda_{2N'}$ must be rotated by an angle of $-\theta=-\pi/4$. Secondly, the rotated charge lattice should be rescaled to accomplish a grid scale of one. We denote these operations as $\CR_{-\theta}$ and $\CM$, respectively.
Under these operations, indeed, the basis vectors $\vec{\ell}_{1}$ and $\vec{\ell}_{2}$ are transformed to $(1, 0)^{\text{T}}$ and $(0, 1)^{\text{T}}$, respectively;
\begin{align}
\begin{aligned}
    \CM\CR_{-\theta}\ :\ \vec{\ell}_{1} \overset{\CR_{-\theta}}{\xlongrightarrow[]{\hspace{2em}}}\ R_{-\theta}\, \vec{\ell}_{1}=(\ell_{1}, 0)^{\text{T}} \ \overset{\CM}{\xlongrightarrow[]{\hspace{2em}}}\ (M^{-1} R_{-\theta})\,  \vec{\ell}_{1}=(1, 0)^{\text{T}} \ , \\
    \CM\CR_{-\theta}\ :\ \vec{\ell}_{2} \overset{\CR_{-\theta}}{\xlongrightarrow[]{\hspace{2em}}}\ R_{-\theta}\, \vec{\ell}_{2}=(0, \ell_{2})^{\text{T}} \ \overset{\CM}{\xlongrightarrow[]{\hspace{2em}}}\ (M^{-1} R_{-\theta})\,  \vec{\ell}_{2}=(0, 1)^{\text{T}} \ ,  
\end{aligned}
\end{align}
 where $\ell_{1}\equiv |\vec{\ell}_{1}|$, $\ell_{2}\equiv |\vec{\ell}_{2}|$, and the matrices $R_{-\theta}$ and $M$ are defined as follows;
\begin{align}\label{eq: rotation and rescaling matrix}
    R_{-\theta}\equiv \begin{pmatrix}
   \cos \theta & \sin \theta \\
   -\sin \theta & \cos \theta 
\end{pmatrix}\qquad , \qquad 
M\equiv \begin{pmatrix}
   \ell_{1} & 0 \\
   0 & \ell_{2} 
\end{pmatrix}\ . 
\end{align}
Therefore, we have succeeded in bringing the charge lattice $\Lambda_{2N'}$ to the original one $\BZ\times \BZ$;
\begin{align}
\CM\CR_{-\theta}\, \Lambda_{2N'}\, \cong\,  \BZ\times \BZ\ . 
\end{align}
We refer the reader to consult the Figure \ref{fig: transition of charged lattice intro}, where the above operations from gauging to rotation to rescaling are illustrated in the case of $c=2$ bosonic torus CFT. Finally, we notice that the charge matrix $K$ defined by \eqref{eq: charge matrix} can be written in terms of the rotation and rescaling matrices;
\begin{align}\label{eq: K matrix}
    K=R_{\, \theta}\, M\ . 
\end{align}

\subsection{Duality}
While the charge lattice indeed comes back to the original one with the above steps, the theory has not yet been restored to the original one $\CT_{g}$. This is because the coupling constants $g'$ of the theory $\CT'_{g'}$ typically differ from the original ones $g$. We, however, can make use of the duality transformation, which maps the theory $\CT'_{g'}$ to the dual one $\widehat{\CT'}_{\widehat{g}\,'}\cong\CT_{g}$, and connects the value of couplings $g'$ to the dual one $\widehat{g}\, '$. Hence, by tuning the original couplings $g$ such that they satisfy the following self-duality condition;
\begin{align}\label{eq: self dual condition}
    \widehat{g}\, '=g\ , 
\end{align}
then the dual theory $\widehat{\CT'}_{\widehat{g}'}$ becomes equivalent to the original one $\CT_{g}$\footnote{Here, we assume that the charge lattice $\widehat{\BZ\times \BZ}$ of the dual theory $\widehat{\CT'}_{\widehat{g}\,'}$ is equivalent to the original one $\BZ\times \BZ$. Indeed, all examples treated in this paper satisfy this property.}. Dualities, of course,  depend on theories, e.g., T-duality for the $c=2$ bosonic torus CFT and electric-magnetic duality for the pure U$(1)\times$U(1) gauge theory. Hence, we relegate the details of self-duality conditions to the subsequent sections.

Following the all above steps, we can conclude that the theory $\CT_g$ is invariant under the diagonal gauging:\,$\CT_{g}/(\BZ^{[q]}_{2N'})_{\text{diag}} \cong \CT_{g}$. Then, we anticipate some (generalized) symmetry associated with the diagonal gauging, which will later be identified with the {\it non-invertible} one.

\section{Example in two dimensions: $c=2$ bosonic torus CFT}
\label{sec:Example in two dimensions: $c=2$ bosonic torus CFT}
In this section, we explore the non-invertible symmetries in the $c=2$ bosonic torus CFT following the method described in section \ref{sec: general theory}. This theory can be described by the following action;
\begin{align}\label{eq: action two compact bosons}
    S[\phi^{1}, \phi^{2}]=\frac{1}{4\pi}\int_{X_{2}}\,G_{IJ}\,  d\phi^{I}\wedge\star \, d\phi^{J}\ \qquad , \qquad I, J=1\, , 2\ ,  
\end{align}
where $X_2$ is the orientable two-dimensional manifold\footnote{Throughout this paper, we only concern the Euclidean spacetime.}, and $\phi^{I}$ is the compact boson with the periodicity $2\pi$;
\begin{align}
    \phi^{I}\sim \phi^{I}+2\pi\ .
\end{align}
Also, $G_{IJ}$ is the real kinetic matrix, which must satisfy the following stability condition;
\begin{align}\label{eq: stability condition}
    G_{11}>0\quad , \quad G_{22}>0\quad , \quad \det G>0\ . 
\end{align}
In this paper, we consider a particular element of the T-duality group O$(2,2,\BZ)$, which maps the kinetic matrix to its inverse;
\begin{align}\label{eq: T-duality}
    \text{T-duality}\ : \ G\mapsto G^{-1} \ .
\end{align}

\noindent The global symmetry of the $c=2$ bosonic torus CFT contains the shift symmetry $\text{U}(1)^{\text{shift}}_{1}\times \text{U}(1)^{\text{shift}}_{2}$, which acts on the vertex operator $V_{\vec{n}}^{[0]}\equiv e^{\i \vec{n}\cdot\vec{\phi}}$ as follows;
\begin{align}
    \text{U}(1)^{\text{shift}}_{1}\times \text{U}(1)^{\text{shift}}_{2}\ : \ V_{\vec{n}}^{[0]}\mapsto e^{\i \vec{\theta}\cdot \vec{n}}\, V_{\vec{n}}^{[0]}\ .
\end{align}
As a first step toward realizing our program \eqref{eq: strategy for constuction}, we perform gauging the diagonal subgroup $(\BZ^{[0]}_{2N'})_{\text{diag}}\subset \text{U}(1)^{\text{shift}}_{1}\times \text{U}(1)^{\text{shift}}_{2}$. Then, the original charge lattice $\BZ\times\BZ$ is reduced to the sublattice defined by \eqref{eq: charge lattice gauging condition};
\begin{align}
    \BZ\times\BZ\longrightarrow \Lambda_{2N'}\ . 
\end{align}
As explained in section \ref{sec: general theory}, we can bring the charge lattice $\Lambda_{2N'}$ to the original one by performing the rotation and rescaling successively. All we need is to perform the duality transformation, hence we proceed to give details on it below. 
\subsection{Self-duality condition}\label{subsubsec: self-duality condition}
In this subsection, we first derive the kinetic matrix after a series of operations (diagonal gauging, rotation, and rescaling), which typically differs from the original one. Next, we show that the T-duality manifestly makes the diagonal gauging $(\BZ_{2N'}^{[0]})_{\text{diag}}$ a symmetry when the kinetic matrix $G_{IJ}$ satisfies the self-duality condition;
\begin{align}\label{eq: self-duality condition G}
    G=K^{\text{T}} G^{-1} K \ .
\end{align}

First of all, under the diagonal gauging and the rotation $\CR_{-\theta}$, the compact boson $\vec{\phi}$ is transformed as follows;
\begin{align}
    \vec{\phi}\mapsto \vec{\phi}'=R_{-\theta}\, \vec{\phi}\ , 
\end{align}
where $R_{-\theta}$ is defined by \eqref{eq: rotation and rescaling matrix}, and the periodicity condition for $\vec{\phi}'$ reads;
\begin{align}\label{eq: periodicity condition}
    \phi'^{1}\sim \phi'^{1}+\frac{2\pi}{\ell_1}\qquad , \qquad \phi'^{2}\sim \phi'^{2}+\frac{2\pi}{\ell_2}\ . 
\end{align}
We should note that the periodicity of $\vec{\phi}'$ is not $2\pi$, and this corresponds to the fact that the grid scale of the charged lattice $\Lambda_{2N'}$ is not one. (See also the flow from the upper left lattice to the upper right one to the bottom middle one in Figure \ref{fig: transition of charged lattice intro}.) Therefore, to restore the original theory, we must make the periodicity of the compact boson $2\pi$. This can be done by the following rescaling transformation;
\begin{align}\label{eq: field redef}
\begin{aligned}
    \vec{\phi}\, ' \mapsto \vec{\phi}''= M\, \vec{\phi}' \ ,  
\end{aligned}
\end{align}
where the matrix $M$ is the rescaling matrix defined in \eqref{eq: rotation and rescaling matrix}. We should notice that the periodicity of $\vec{\phi}''$ is $2\pi$, and this restoration can be seen in the flow from the bottom middle lattice to the upper left one in Figure \ref{fig: transition of charged lattice intro}. In summary, the diagonal gauging $(\BZ_{2N'}^{[0]})_{\text{diag}}$, rotation and rescaling transform the compact boson field $\vec{\phi}$ as follows;
\begin{align}\label{eq: transformation law under gauging and K}
\begin{aligned}
        \vec{\phi}\mapsto \vec{\phi}''
        &=MR_{-\theta}\, \vec{\phi}\  \\
        &=K^{\text{T}} \vec{\phi}\ , 
\end{aligned}
\end{align}
where we used \eqref{eq: K matrix}. Since the periodicities of both compact bosons $\vec{\phi}$ and $\vec{\phi}''$ are $2\pi$, the above map \eqref{eq: transformation law under gauging and K} can be rephrased by the transformation law for the kinetic matrix $G_{IJ}$; 
\begin{align}
    G \, \mapsto\,  K^{-1}\,  G \, (K^{\text{T}})^{-1} \ . 
\end{align}

\noindent As stressed in earlier times, the kinetic matrix after the series of operations, takes the different values from the original one. However, by making full use of the T-duality, we can put the theory back to the original one. The T-duality transformation \eqref{eq: T-duality}, indeed, maps the deformed kinetic matrix $K^{-1} G (K^{\text{T}})^{-1}$ to its inverse;
\begin{align}
    \text{T-duality : } K^{-1} G (K^{\text{T}})^{-1}\mapsto K^{\text{T}} G^{-1} K \ . 
\end{align}
Therefore, if we choose the kinetic matrix $G_{IJ}$ such that it satisfies the following self-duality condition;
\begin{align}\label{eq: self-duality main text}
    G=K^{\text{T}} G^{-1} K\ , 
\end{align}
the diagonal gauging $(\BZ_{2N'}^{[0]})_{\text{diag}}$ becomes a true symmetry. In the following, we denote the solution to \eqref{eq: self-duality main text} by $G^{*}$. 
By noticing $\det G=2N'$, we easily obtain the solution to the self-duality condition;
\begin{align}\label{eq: self-duality solution}
    G^{*} =\sqrt{-\frac{2N'}{D}}
    \begin{pmatrix}
        2K_{11} & K_{12}+K_{21} \\
   K_{12}+K_{21} & 2K_{22}
    \end{pmatrix}\ , 
\end{align}
where $D$ is defined by
\begin{align}\label{eq: discriminatnt}
    D\equiv (K_{12}+K_{21})^{2}-4K_{11}K_{22}\ . 
\end{align}
We should note that the kinetic matrix $G_{IJ}$ must be real in physical theory, hence $D$ takes the negative value;
\begin{align}
    D<0\ . 
\end{align}
Our particularly interest in this paper is the diagonal gauging $(\BZ_{2N'}^{[0]})_{\text{diag}}$, hence the above physical condition restricts the values of $N'$ as follows:
\begin{align}
    N'=1\, , 2\, , 3\, , 4\, , 5\, .
\end{align}
Interestingly, the self-dual solution \eqref{eq: self-duality solution} is the same one known as {\it complex multiplication (CM) point} which is a more generic point than a RCFT in the $c=2$ bosonic torus CFT \footnote{We thank Justin Kaidi for plentiful discussions on this point.} \cite{Gukov:2002nw}. To see this, we put the kinetic matrix into the complex structure modulus $\tau$ 
defined by;
\begin{align}\label{eq: complex structure modulus}
    \tau\equiv \frac{G_{12}}{G_{22}}+\i \frac{\sqrt{\det G}}{G_{22}}\ , 
\end{align}
then the complex structure modulus at the self-dual point $\tau^{*}$ satisfies the following quadratic equation;
\begin{align}\label{eq: quadratic equation}
    K_{22}\, (\tau^{*})^{2}-(K_{12}+K_{21})\tau^{*}+K_{11}=0\ . 
\end{align}
Importantly, the discriminat of the above quadratic equation \eqref{eq: quadratic equation} is precisely same as $D$ defined in \eqref{eq: discriminatnt}, and its negative property $D<0$ ensures that $\tau^*$ belongs to the imaginary quadratic number field $\BQ(\sqrt{D})$ \cite{Gukov:2002nw};
\begin{align}\label{eq: imaginary quadratic number property}
    \tau^* \in \BQ(\sqrt{D})\ . 
\end{align}
Since it is known that the elliptic curves with the modular parameter $\tau^*$ satisfying \eqref{eq: imaginary quadratic number property} have complex multiplication properties, such modulus is called the CM point. 

Then, the following natural question arises; When can the CM point be lifted up to the RCFT? We can show that only if the charge matrix $K$ is symmetric, namely $K^{\text{T}}=K$, this promoting occurs.
\noindent The proof is as follows. In order for this lifting to be achieved, it is sufficient to show that the complexified K\"{a}hler modulus $\rho$ also needs to belong to the {\it same} imaginary quadratic number field $\BQ(\sqrt{D})$ \cite{Gukov:2002nw}. Now, the complexified K\"{a}hler modulus $\rho$ is given by the pure imaginary number due to the absence of the B-field;
\begin{align}\label{eq: Kahler moduli}
    \rho\equiv \i \sqrt{\det G}\ , 
\end{align}
and at the self-duality point, the modulus $\rho$ becomes $\rho^* =\i \sqrt{2N'}$. Hence, when there exist the integers $\alpha, \beta$ and $\gamma$ such that;
\begin{align}
    \alpha (\rho^{*})^2 +\beta \rho^{*} +\gamma=0\qquad \text{and} \qquad \beta^2 -4\alpha\gamma=D\ , 
\end{align}
the CM points can get promoted to RCFT ones. We first notice that $\beta=0$ due to the pure imaginary property of $\rho^*$, then resulting in $\gamma=2N'\alpha$. Next, we can rewrite the discriminant $D$ given in \eqref{eq: discriminatnt} as follows;
\begin{align}
    D=(K_{12}-K_{21})^2 -8N'\ , 
\end{align}
where the formula $\det K =2N'$ is used. Therefore, to realize $\rho^* \in \BQ(\sqrt{D})$, there must exist some integers $\alpha$, $k$ such that
\begin{align}\label{eq: alpha eq}
    \alpha^2=k^{2}\left[1-\frac{(K_{12}-K_{21})^2}{8N'}\right]\ . 
\end{align}
Note that we cannot find some nice integers $\alpha$ and $k$ such that they are solutions of the above equation \eqref{eq: alpha eq} in the case of the diagonal gauging\footnote{One may notice we can make the charge matrix $K$ be symmetric if we exchange the two basis vectors $\vec{\ell}_{1}$ and $\vec{\ell}_{2}$. In that case, however, we can never obtain the integer $\alpha$ because of $\det K=-2N'$.}, as can be easily checked from \eqref{eq: charge matrix}. (Recall that the possible values of $N'$ are the integers ranging from $1$ to $5$.) Hence, we arrive at the following conclusion;
\begin{shaded}
    \noindent A (generalized) symmetry associated with the diagonal gauging $(\BZ^{[0]}_{2N'})_{\text{diag}}$ becomes emergent at the {\it irrational} CFT point.
\end{shaded}

\paragraph{Emergent $\BZ_{2}$ symmetry at the self-dual point.} Finally, we close this subsection by mentioning the non-trivial emergent $\BZ_{2}$ symmetry at the irrational CFT point. We should note that the self-duality condition \eqref{eq: self-duality main text} is invariant under the replacement of the charge matrix $K$ with its transposed one $K^{\text{T}}$;
\begin{align}
    K^{\text{T}} G^{-1} K=G \qquad \Longleftrightarrow \qquad K G^{-1} K^{\text{T}}=G\ . 
\end{align}
This implies some emergent $\BZ_{2}$ symmetry at the self-dual point, and we find out that the mapping $K\mapsto K^{\text{T}}$ can be realized by the transformation of the compact boson field; 
\begin{align}\label{eq:action of charge conjugation}
    \phi^I \mapsto \phi'^{I}=\SlS_{IJ}\, \phi^{J} \qquad , \qquad 
    \SlS=    
    \begin{pmatrix}
    1 & 0 \\
    1-N' & -1 
    \end{pmatrix}
    .
\end{align}
Indeed, the matrix $\SlS$ satisfies the following properties;
\begin{align}
\begin{aligned}\label{eq:properties of S}
        \SlS^2&=1_{2\times 2}\qquad ,\qquad
    \SlS^{\text{T}}\, G^{*}\, \SlS&=G^{*}\qquad , \qquad
    \SlS^{\text{T}}\, K\,  \SlS&=K^{\text{T}} \ ,
\end{aligned}
\end{align}
and we can easily check that the theory at the self-dual point is invariant under the $\BZ_{2}$ transformation \eqref{eq:action of charge conjugation}. This emergent $\BZ_2$ symmetry plays a crucial role in discussing the fusion algebra, and we denote the topological defect associated with this emergent $\BZ_2$ symmetry by $\CS$.

\subsection{Duality defect}\label{subsec: Duality defect}
In this subsection, following the spirit of \cite{Choi:2021kmx, Kaidi:2021xfk}, we derive the duality defect associated with the diagonal gauging $(\BZ_{2N'}^{[0]})_{\text{diag}}$. First of all, we divide the ambient spacetime into the left and right regions separated by the co-dimension one defect residing at $x=0$. Then, we propose that the duality defect $\CD$ can be expressed by the following Lagrangian;
\begin{align}\label{eq: duality defect in 2d}
    \CD \ : \ \frac{\i}{2\pi}\int_{x=0} K_{IJ}\, \phi_{\text{L}}^{I}\, d \phi_{\text{R}}^{J} \ . 
\end{align}
where $\phi_{\text{L}}$ and $\phi_{\text{R}}$ are the compact boson fields which are located in the bulks $x<0$ and $x>0$, respectively. We should note that the duality defect \eqref{eq: duality defect in 2d} is gauge-invariant since the charge matrix $K$ is an integer matrix. In the following, we explicitly show that only when the bulk kinetic matrix is tuned to be self-dual one $G^{*}$, the duality defect $\CD$ correctly reflects the sequence of operations in \eqref{eq: strategy for constuction} by seeing the boundary conditions of the left and right fields. Finally, we give some comments on the topological property of the duality defect and its orientation-reversion. 
\begin{figure}[t]
    \centering
 \begin{tikzpicture}[scale=1]
\draw[black!70, very thick] (0, 0)--(10, 0)--(10, 5)--(0, 5)--(0, 0);
\draw[red!70, very thick] (5, 0) node[below, black]{$x=0$}--(5, 5) node[above]{$\CD : \frac{\i}{2\pi}\int_{x=0}K_{IJ}\, \phi_{\text{L}}^{I}d\phi_{\text{R}}^{J}$}; 
\node[black] at (2.5, 2.5) {$\frac{1}{4\pi}\int_{\text{L}} G^{*}_{IJ}\, d\phi_{\text{L}}^{I}\, \wedge \star d\phi_{\text{L}}^{J}$}; 
\node[black] at (7.5, 2.5) {$\frac{1}{4\pi}\int_{\text{R}} G_{IJ}^{*}\,  d\phi_{\text{R}}^{I}\, \wedge \star d\phi_{\text{R}}^J$}; 
  \end{tikzpicture}
    \caption{Pictorical representation of the duality defect $\CD$.}
    \label{fig: half-space gauging pictorical representation}
\end{figure}

The combined system of the bulk theory and the duality defect can be described in the following action (see Figure \ref{fig: half-space gauging pictorical representation}.); 
\begin{align}\label{eq: duality line}
    \frac{1}{4\pi}\int_{\text{L}}G^{*}_{IJ}\, d\phi^{I}_{\text{L}}\wedge\star \, d\phi^{J}_{\text{L}}+\frac{1}{4\pi}\int_{\text{R}}G^{*}_{IJ}\, d\phi^{I}_{\text{R}}\wedge\star \, d\phi^{J}_{\text{R}}+\frac{\i}{2\pi}\int_{x=0} K_{IJ}\, \phi_{\text{L}}^{I}\, d \phi_{\text{R}}^{J}\ , 
\end{align}
then the variations of the left and right compact boson fields give rise to the following boundary conditions at $x=0$;
\begin{align}\label{eq: boundary condition 1}
    x=0\ &: \ \i\,  G^{*}_{IJ}\star d\phi^{J}_{\text{L}}= K_{IJ}\, d\phi_{\text{R}}^{J}\ , 
\end{align}
\begin{align}\label{eq: boundary condition 2}
    x=0\ &: \ \i\,  G^{*}_{IJ}\star d\phi^{J}_{\text{R}}= K_{JI}\, d\phi_{\text{L}}^{J}\ . 
\end{align}
We can readily check the equivalence of these two conditions by using the self-duality condition \eqref{eq: self-duality main text}. Concretely speaking, we can obtain the latter boundary condition \eqref{eq: boundary condition 2} from the former one \eqref{eq: boundary condition 1} by acting the Hodge dual operation $\star$ on both hand sides in \eqref{eq: boundary condition 1}, and using the self-duality condition \eqref{eq: self-duality main text}, and vice versa. Furthermore, we should note that by rewriting the matrix $K$ in terms of the rotation matrix and the rescaling one, the boundary condition becomes;
\begin{align}
    x=0\ &: \ \i\, G^{*}_{IJ}\, d\phi_{\text{R}}^{J}= \star\left((M R_{-\theta})_{IJ}\,  d\phi^{J}_{\text{L}}\right)\ .
\end{align}
This can be interpreted as first performing a $(\BZ^{[0]}_{2N'})_{\text{diag}}$ gauging to rotate the compact boson by angle $-\theta$ to rescale by the matrix $M$, and finally performing the T-duality transformation. This observation corroborates that our construction \eqref{eq: strategy for constuction} can be realized by insertion of the duality defect $\CD$ defined by \eqref{eq: duality defect in 2d} into the spacetime.

In addition, we insist that the duality defect $\CD$ becomes topological when $G=G^{*}$. Although this is clear from the viewpoint of the half-space gauging \cite{Choi:2021kmx, Kaidi:2021xfk}, we provide another proof in the spirit of \cite{Kapustin:2009av,Niro:2022ctq}. To show this topological property, it is enough to show that the energy-momentum tensors must satisfy the following matching condition;
\begin{align}\label{eq: matching condition}
   x=0\ &: \ n_{\mu}(T^{\mu\nu}_{\text{L}}-T^{\mu\nu}_{\text{R}})=0 \ . 
\end{align}
Here, $T^{\mu\nu}_{\text{L}}$ and $T^{\mu\nu}_{\text{R}}$ are the energy-momentum tensors in the left and right regions, and given by
\begin{align}
    \begin{aligned}
    T^{\mu\nu}_{\text{L}}&= \frac{1}{4\pi}G_{IJ}\, \partial_{\alpha}\phi^{I}_{\text{L}}\partial_{\beta}\phi^{J}_{\text{L}}\left(\frac{1}{2}\delta^{\alpha\beta}\delta^{\mu\nu}-\delta^{\mu\alpha}\delta^{\nu\beta}\right)\ , \\
        T^{\mu\nu}_{\text{R}}&= \frac{1}{4\pi}G_{IJ}\, \partial_{\alpha}\phi^{I}_{\text{R}}\partial_{\beta}\phi^{J}_{\text{R}}\left(\frac{1}{2}\delta^{\alpha\beta}\delta^{\mu\nu}-\delta^{\mu\alpha}\delta^{\nu\beta}\right)\ , 
    \end{aligned}
\end{align}
respectively, and $n^{\mu}$ is the normal vector to the duality defect $\CD$. We can easily prove that the matching condition \eqref{eq: matching condition} can be achieved by using \eqref{eq: self-duality main text} and \eqref{eq: boundary condition 1}.
As a result of this reasoning, we can conclude that the duality defect $\CD$ is topological.

Finally, we comment on the orientation-reversing of the duality defect $\CD$. Its orientation reversal $\overline{\CD}(M)$ is defined by \cite{Roumpedakis:2022aik, Choi:2022zal};
\begin{align}
    \overline{\CD}(M)=\CD(\overline{M})\ , 
\end{align}
where $M$ is the support manifold of the duality defect, namely $x=0$, and $\overline{M}$ denotes the orientation reversal of $M$.
The defect action of $\overline{\CD}$ can be obtained by swapping $\phi_L$ with $\phi_R$, and flipping the overall sign stemming from the orientation-reversion of $M$;
\begin{align}
    \overline{\Duality}(M)\ : \ \frac{\i}{2\pi}\int_{x=0} K_{JI}\, \phi_{\text{L}}^{I}\, d \phi_{\text{R}}^{J} \ . 
\end{align}
Note that we can also obtain $\overline{\Duality}$ only by replacing the charge matrix $K$ with its transposed one in the duality defect $\CD$. We can realize this replacement by utilizing the emergent $\BZ_2$ symmetry discussed in section \ref{subsubsec: self-duality condition}, and write the orientation-reversed duality defect $\overline{\CD}$ in terms of $\CD$ and $\CS$;
\begin{align}\label{eq:barduality relation}
    \overline{\Duality}=\CS\times\Duality\times\CS\,.
\end{align}
We should notice that the orientation-reversed duality defect $\overline{\Duality}$ can be interpreted as the duality defect obtained by gauging $\BZ_{2N'}^{[0]}$ symmetry generated by $\etaSp$;
\begin{align}\label{eq:relation of etaSp}
    \etaSp=\CS\times\etap\times\CS\,.
\end{align}
This is because, after gauging this $\BZ_{2N'}^{[0]}$ symmetry, the charge matrix is given by the transposed matrix $K^{\text{T}}$.

\subsection{Non-commutative fusion algebra}
\label{sec:Fusion rules}
In this subsection, we describe various fusion rules involving the duality defect $\CD$ introduced in section \ref{subsec: Duality defect}. Since the derivations of the fusion algebra require somewhat technical calculations, we just digest our results here.  
If the readers have some interest in the detailed calculations, we refer to reading appendix \ref{sec: derivation of fusion algebra}, where some skipped derivations are demonstrated.

First of all, we consider the fusion rules between the duality defect $\CD$ and the $(\BZ_{2N'}^{[0]})_{\text{diag}}$ shift symmetry generator $\eta_{\vec{p}}\, (\Sigma)$ defined by;
\begin{align}\label{eq: def of shift generator}
    \eta_{\vec{p}}\, (\Sigma)\equiv \exp\left[-\frac{p^{I}}{2N'}\int_{\Sigma}G_{IJ}\star d\phi^{J}\right]\qquad , \qquad \vec{p}\equiv (1,1)^{\text{T}}\ ,
\end{align}
where $\Sigma$ is the parallel line to the duality defect $\CD$. Interestingly, unlike the ordinary fusion rules of the duality defect, $\eta_{\vec{p}}\times \CD$ and $\CD\times \eta_{\vec{p}}$ do not give rise to the same results in general. If we bring the $(\BZ_{2N'}^{[0]})_{\text{diag}}$ shift symmetry generator $\eta_{\vec{p}}$ closer to the duality defect $\CD$ from the left side, the fusion rule $\eta_{\vec{p}}\times \CD$ reads; 
\begin{align}\label{eq: eta times D}
    \eta_{\vec{p}}\times \CD\ : \ \frac{\i}{2\pi}\int_{x=0}\,K_{IJ}\, \phi^{I}_{\text{L}}\, d \phi_{\text{R}}^{J}
             -\frac{\i p^{I}K_{IJ}}{2N'}\int_{x=0}\,  d\phi^J_{\text{R}}\ . 
\end{align}
Here, we should recall $p^{I}K_{IJ}=0 \mod 2N'$, hence the last term in \eqref{eq: eta times D} becomes trivial and can be dropped. This implies that the symmetry generator $\eta_{\vec{p}}$ is absorbed into the duality defect $\CD$, and the fusion rule $\eta_{\vec{p}}\times \CD$ becomes as follows;
\begin{align}\label{eq: eta times D 2}
    \eta_{\vec{p}}\times \CD=\CD\ . 
\end{align}
From this, it turns out that the duality defect $\Duality$ is non-invertible\footnote{This can be readily checked as follows. Suppose that the duality defect $\CD$ is invertible, i.e., $\CD\times \CD^{-1}=1$, this contradicts with the fusion rule derived in \eqref{eq: eta times D 2}; 
\begin{align}
\Duality\times\Duality^{-1}=\eta_{\vec{p}}\times \CD \times \CD^{-1}=\eta_{\vec{p}}\neq1\ . 
\end{align}}.
On the other hand, if we put the $(\BZ_{2N'}^{[0]})_{\text{diag}}$ shift symmetry generator $\eta_{\vec{p}}$ to the duality defect $\CD$ from the right, the {\it non-trivial} $\BZ_{2N'}^{[0]}$ winding symmetry generator becomes emergent on the left side of the duality defect;
\begin{align}\label{eq: D times eta}
    \CD\times \eta_{\vec{p}}\ : \ \frac{\i}{2\pi}\int_{x=0}\,K_{IJ}\, \phi^{I}_{\text{L}}\, d \phi_{\text{R}}^{J}
        -\frac{\i p^{J}K_{IJ}}{2N'}\int_{x=0}\,  d\phi^I_{\text{L}}\, , 
\end{align}
since $p^{J}K_{IJ}\not=0 \mod 2N'$ in general. We denote this emergent $\BZ^{[0]}_{2N'}$ winding symmetry generator by $\widetilde{\eta}_{K\vec{p}}$, then the fusion rule $\CD \times \eta_{\vec{p}}$ becomes as follows;
\begin{align}\label{eq: D times eta 2}
    \CD \times \eta_{\vec{p}} = \widetilde{\eta}_{K\vec{p}} \times \CD\ . 
\end{align}
By comparing the above results \eqref{eq: eta times D 2} and \eqref{eq: D times eta 2}, we can conclude that the obtained fusion rules are non-commutative, namely $\eta_{\vec{p}}\times \CD\not=\CD\times \eta_{\vec{p}}$. What happens if we close the $\BZ^{[0]}_{2N'}$ winding symmetry generator $\widetilde{\eta}_{K\vec{p}}$ to the right side of the duality defect $\CD$\,? The fusion rule $\CD\times\widetilde{\eta}_{K\vec{p}}$ can be calculated as follows;
\begin{align}\label{eq: D times wind}
    \CD\times \widetilde{\eta}_{K\vec{p}}\ : \ \frac{\i}{2\pi}\int_{x=0}\,K_{IJ}\, \phi^{I}_{\text{L}}\, d \phi_{\text{R}}^{J}
        +\frac{(\vec{p}^{\,\text{T}}\,K^{\text{T}}\, K^{-1})^{I}}{2N'}\int_{x=0}\,  G_{IJ}\,\star d\phi^I_{\text{L}}\, .
\end{align}
From \eqref{eq: def of shift generator}, the last term in \eqref{eq: D times wind} is none other than the shift symmetry generator $\eta_{MK\vec{p}}$\,;
\begin{align}
    \eta_{MK\vec{p}}\, (\Sigma)\equiv \exp\left[-\frac{(\vec{p}^{\,\text{T}}K^{\text{T}}M^{\text{T}})^{I}}{2N'}\int_{\Sigma}G_{IJ}\star d\phi^{J}\right] \qquad , \qquad M\equiv (K^{\text{T}})^{-1} \ , 
\end{align}
and the fusion rule $\CD\times \widetilde{\eta}_{K\vec{p}}$ becomes as follows;
\begin{align}
    \CD\times \widetilde{\eta}_{K\vec{p}}=\eta_{MK\vec{p}}\times \CD\ .
\end{align}
We should notice that $\eta_{MK\vec{p}}$ is the symmetry generator associated with $\BZ^{[0]}_{(2N')^{2}}$ shift symmetry. If we moreover put this the shift symmetry generator $\eta_{MK\vec{p}}$ to $\CD$ from the right, $\BZ^{[0]}_{(2N')^{2}}$ winding symmetry generator $\widetilde{\eta}_{KMK\vec{p}}$ appears in the left side;
\begin{align}
    \CD\times \eta_{MK\vec{p}}=\widetilde{\eta}_{KMK\vec{p}}\times \CD\ . 
\end{align}
We can straightforwardly keep going the above discussions, and eventually obtain the following fusion rules:
\begin{align}
    \CD\times \eta_{(MK)^{i}\,\vec{p}}=\widetilde{\eta}_{K(MK)^{i}\,\vec{p}}\times \CD\quad , \quad \CD\times \widetilde{\eta}_{K(MK)^{i}\,\vec{p}}=\eta_{(MK)^{i+1}\,\vec{p}}\times \CD\quad , \quad i=0, 1, 2, \cdots\ ,  
\end{align}
where $\eta_{(MK)^{i}\,\vec{p}}$ and $\widetilde{\eta}_{K(MK)^{i}\,\vec{p}}$ are $\BZ^{[0]}_{(2N')^{i}}$ shift and winding symmetry generators, respectively. This implies that the fusion algebra concerning the non-invertible duality defect constructed from the diagonal gauging is \emph{infinitely} generated. Notably, this is consistent with the general property of the irrational CFT where the number of topological defect lines is expected to be infinite.

Interestingly, we find out the closed fusion subalgbera of the infinitely generated fusion algebra described above. To see this, we first redefine the duality defect $\CD$ by dressing the $\BZ_{2N'}^{[0]}$ winding symmetry generator;
\begin{align}\label{eq: dressed duality defect}
    \Duality_{s}(\Sigma)\ : \  \int_{\Sigma}\,\left(\frac{\i}{2\pi}K_{IJ}\, \phi^{I}_{\text{L}}\, d \phi_{\text{R}}^{J}
    -\frac{\i s\,  p^{J}K_{IJ}}{2N'}\,  d\phi^I_{\text{L}}\right) \ ,
\end{align}
which is labelled by the $\BZ_{2N'}$ element $s=0, 1, \cdots 2N'-1$. The fusion rules between the dressed duality defect $\CD_{s}$ and the $(\BZ_{2N'}^{[0]})_{\text{diag}}$ shift symmetry generator $\eta_{\vec{p}}$ becomes;
\begin{align}
    &\eta_{\vec{p}}\times\Duality_{s}=\Duality_{s}\label{eq:e times Ds}\ ,\\
    &\Duality_{s}\times\eta_{\vec{p}}=\Duality_{s+1}\ .\label{eq:Ds times e}
\end{align}
However, as it is, the fusion algebra is not closed, which can be seen from the direct calculation of $\CD_{s_1}\times \CD_{s_2}$;
\begin{align}
    \CD_{s_1}\times \CD_{s_2}\ : \ \frac{\i}{2\pi}\int_{x=0}(K_{IJ}\phi^{I}_{\text{L}}-K_{JI}\phi^{I}_{\text{R}})d\phi^{J}_{\text{M}}-\frac{\i s_{2}\,  p^{J}K_{IJ}}{2N'}\int_{x=0}d\phi^{I}_{\text{M}}\ . 
\end{align}
To our best effort, we cannot write the above result in a closed form by using the known topological defects.
Hence, further modification for the duality defect is needed to close the fusion algebra. After some trial and error, we find out that the following combination works well for the closure of the fusion algebra;\footnote{We note that orientation-reversed dressed duality defect $\BarhatDuality{s_1,s_2}(M)\equiv\hatDuality{s_1,s_2}(\overline{M})$ can be expressed in terms of the dressed duality defect $\widehat{\CD}_{s_1 , s_2}$ as follows;
\begin{align}
\begin{aligned}
        \BarhatDuality{s_1,s_2}
    &=\etaSp^{-s_2}\times\SlS\times\overline{\Duality}\times\etaSp^{s_1}\\
    &=\etaSp^{-s_2}\times\Duality\times\SlS\times\etaSp^{s_1}\\
    &=\hatDuality{-s_2,-s_1}\,.
\end{aligned}
\end{align}} 
\begin{align}\label{eq: dressed duality defect}
    \hatDuality{s_1,s_2}\equiv (\etaSp)^{s_1}\times\CD_{s_2}\times \CS\quad , \quad s_1 , s_2 =0,\cdots, 2N'-1\ , 
\end{align}
and we arrive at the following conclusion;
\begin{shaded}
    \noindent \textbf{\mylabel{result}{Non-commutative fusion subalgebra at the irrational CFT point}.} Non-invertible symmetry constructed from the half-space gauging of $(\BZ_{2N'}^{[0]})_{\text{diag}}$ becomes emergent at the {\it irrational} CFT point. The fusion algebra is {\it non-commutative} and {\it infinitely} generated. We find out the closed fusion subalgebra;
    \begin{align}\label{eq: fusion category}
        \begin{aligned}
        &\hatDuality{s_1,s_2}\times \hatDuality{s_3,s_4}=\widehat{\CC}_{s_2+s_3,s_1+s_4}\ ,\\
      &\hatDuality{s_1,s_2}\times\eta_{\vec{p}}=\eta_{\vec{p}}\times\hatDuality{s_1,s_2}=\hatDuality{s_1,s_2}\ ,\\
      & \eta_{\vec{p}}\times \widehat{\CC}_{s_1,s_2}=\widehat{\CC}_{s_1,s_2}\times \eta_{\vec{p}}=\widehat{\CC}_{s_1,s_2}\ , \\
      &\widehat{\CD}_{s_1,s_2}\times \widehat{\CC}_{s_3,s_4}=2N'\, \hatDuality{s_1+s_3,s_2+s_4} \ , \\ &\widehat{\CC}_{s_1,s_2}\times \widehat{\CD}_{s_3,s_4}=2N'\, \hatDuality{s_2+s_3,s_1+s_4}\ , \\
      &\widehat{\CC}_{s_1,s_2}\times\widehat{\CC}_{s_3,s_4}=2N'\,\widehat{\CC}_{s_1+s_3,s_2+s_4}\ ,\\
        &\eta_{\vec{p}}^{2N'}=1\ ,
        \end{aligned}
    \end{align}
    where $\widehat{\CC}_{s_1,s_2}$ is the projection operator associated with $(\BZ_{2N'}^{[0]})_{\text{diag}}$ shift symmetry, dressed with the $\BZ_{2N'}^{[0]}$ winding symmetry generator;
    \begin{align}\label{eq: projection operator}
    \begin{aligned}
        \widehat{\CC}_{s_1,s_2}(\Sigma)\equiv &\exp\left[-\frac{\i s_1 p^{J} K_{IJ}}{2N'}\int_{\Sigma}d\phi_{\text{L}}^{I}\right]\, 
         \\
        &\times\int \CD\varphi\exp\left[-\frac{\i}{2\pi}\int_{\Sigma} K_{IJ}\left(\phi_{\Left}^I-\phi_{\Right}^I-\frac{2\pi s_{2}}{2N'}\, \SlS^{IK}\, p_{K}\right)d\varphi^J\,\right]\ . 
    \end{aligned}
    \end{align}
 \end{shaded}

\noindent We can easily check that the above fusion algebra satisfies the associativity condition. For instance, the fusion rule $(\hatDuality{s_1,s_2}\times \hatDuality{s_3,s_4})\times \hatDuality{s_5, s_6}$ becomes as follows;
\begin{align}\label{eq: associativity check1}
\begin{aligned}
        (\hatDuality{s_1,s_2}\times \hatDuality{s_3,s_4})\times \hatDuality{s_5, s_6}&=\widehat{\CC}_{s_2 +s_3, s_1 +s_4}\times\hatDuality{s_5, s_6}  \\
        &=2N' \hatDuality{s_1 +s_4 +s_5 , s_2 +s_3 +s_6}\ . 
\end{aligned}
\end{align}
On the other hand, the fusion rule $\hatDuality{s_1,s_2}\times (\hatDuality{s_3,s_4}\times \hatDuality{s_5, s_6})$ can be evaluated as follows;
\begin{align}\label{eq: associativity check2}
    \begin{aligned}
        \hatDuality{s_1,s_2}\times (\hatDuality{s_3,s_4}\times \hatDuality{s_5, s_6}) &= \hatDuality{s_1,s_2}\times \widehat{\CC}_{s_4 +s_5 , s_3 +s_6} \\
        &=2N' \hatDuality{s_1 +s_4 +s_5 , s_2 +s_3 +s_6}\ .
    \end{aligned}
\end{align}
From \eqref{eq: associativity check1} and \eqref{eq: associativity check2}, the associativity condition does hold;
\begin{align}
    (\hatDuality{s_1,s_2}\times \hatDuality{s_3,s_4})\times \hatDuality{s_5, s_6}=\hatDuality{s_1,s_2}\times (\hatDuality{s_3,s_4}\times \hatDuality{s_5, s_6}) \ .
\end{align}
For other defects, we can show the associativity in a similar manner to the above.

Here, it is instructive to compare the above fusion algebra \eqref{eq: fusion category} with the one obtained by gauging the product subgroup $\BZ_{N_1}\times \BZ_{N_2}\subset$U$(1)_{1}\times$U(1)$_{2}$. In this case, the charge matrix is given by the following diagonal form;
\begin{align}
    \begin{aligned}
        K=\begin{pmatrix}
            N_1 & 0 \\
            0 & N_2
        \end{pmatrix}\ , 
    \end{aligned}
\end{align}
then the self-dual kinetic matrix is also diagonal;
\begin{align}\label{eq: self-duality kinetic matrix S2}
    G=\begin{pmatrix}
            N_1 & 0 \\
            0 & N_2
        \end{pmatrix}\ . 
\end{align}
The above result \eqref{eq: self-duality kinetic matrix S2} shows that we can split the $c=2$ bosonic torus CFT into the double $c=1$ compact boson CFTs, whose radiuses are given by $\sqrt{N_1}$ and $\sqrt{N_2}$. By recalling the facts about non-invertible symmetries in $c=1$ compact boson CFT \cite{Choi:2021kmx, Kaidi:2021xfk}, we can convince that the non-invertible symmetry from the gauging $\BZ_{N_1}\times \BZ_{N_2}$ becomes emergent at a RCFT point, and the resulting fusion algebra is given by the Tambara-Yamagami category \eqref{eq: Tambara-Yamagami c=1}. This observation is consistent with the fact that we can undress totally of various global symmetry generators from the dressed duality defect $\widehat{\CD}_{s_1 , s_2}$, and the resulting fusion algebra \eqref{eq: fusion category} is reduced to the Tambara-Yamagami category.

\section{Example in four dimensions: pure U(1)$\times$U(1) gauge theory}
\label{sec: Example in four dimensions}
In this section, we provide a four-dimensional example that has a non-invertible symmetry obtained from the diagonal gauging. In particular, we consider the pure U$(1)\times $U$(1)$ gauge theory; 
\begin{align}\label{eq: action of two gauge theory}
    S[A^{1}, A^{2}]=\frac{1}{4\pi}\int_{M_{4}}\,\CG_{IJ}\,  dA^{I}\wedge\star \, dA^{J}\ \qquad , \qquad I, J=1\, , 2\ ,
\end{align}
where $A^{1}$ and $A^{2}$ are U(1) gauge fields. Note that this model is the higher-dimensional analog of the $c=2$ bosonic torus CFT, and has the same symmetry structures with that. (See Table \ref{table:2 and 4 dimensions}.) Therefore, we can parallelly discuss the non-invertible symmetry in this model based on the method accomplished in section \ref{sec:Example in two dimensions: $c=2$ bosonic torus CFT}.

This theory has the electric one-form symmetry U$(1)^{\text{ele}}_{1}\times$U$(1)^{\text{ele}}_{2}$ \cite{Gaiotto:2014kfa} and the electric-magnetic duality: $\CG\mapsto \CG^{-1}$.
First, we gauge the diagonal subgroup $(\BZ_{2N'}^{[1]})_{\text{diag}}\subset$ U$(1)^{\text{ele}}_{1}\times$U$(1)^{\text{ele}}_{2}$, generated by; 
\begin{align}
    \eta_{\vec{p}}\, (\Sigma_{2})=\exp\left[-\frac{\, p^{I}}{2N'}\int_{\Sigma_{2}} \CG_{IJ}\star dA^{J}\right]\quad , \quad \vec{p}\equiv(1,1)\ , 
\end{align}
where $\Sigma_2$ is a two-dimensional closed manifold.
When the kinetic matrix $\CG_{IJ}$ satisfies the self-duality condition $K^{\text{T}} \CG^{-1} K=\CG$, namely its solution is given by
\begin{align}\label{eq: self-duality condition CG}
    \CG^* = \sqrt{-\frac{2N'}{D}}
    \begin{pmatrix}
        2K_{11} & K_{12}+K_{21} \\
   K_{12}+K_{21} & 2K_{22}
    \end{pmatrix}\quad ,\quad     D=(K_{12}+K_{21})^{2}-4K_{11}K_{22}\ , 
\end{align}
the diagonal gauging $(\BZ^{[1]}_{2N'})_{\text{diag}}$ becomes a non-invertible symmetry. Furthermore, the (topological) duality defect $\Duality$ and its orientation reversal $\overline{\Duality}$ is obtained by the same procedure as discussed in section \ref{sec: general theory}, and its defect action can be written as the following;
\begin{align}\label{eq: duality defect in 2}
    \Duality\ &: \ \frac{\i}{2\pi}\int_{M_3} K_{IJ}\, A_{\text{L}}^{I}\, d A_{\text{R}}^{J} \ ,\\
     \overline{\Duality}\ &: \ -\frac{\i}{2\pi}\int_{M_3} K_{JI}\, A_{\text{L}}^{I}\, d A_{\text{R}}^{J} \ , 
\end{align}
where $M_3$ is a three dimensional manifold, and $A_{\text{L}}$ and $A_{\text{R}}$ are gauge fields living in left and right regions, respectively.
As with the two dimensions \eqref{eq:barduality relation}, we can rewrite $\overline{\Duality}$ in a following way;
\begin{align}
\overline{\Duality}=\CS\times\Duality\times\CS\times\CU\,,
\end{align}
where $\CS$ and $\CU$ are $\BZ_2$ symmetry defects which act on the gauge field $A^{I}$ as follows;
\begin{align}
    \CS \ : \ A^I\mapsto \SlS_{IJ}\,A^J\qquad , \qquad \CU\ : \ A^I\mapsto -A^I\ .
\end{align}
Here, the matrix $\SlS$ is defined in \eqref{eq:action of charge conjugation}. The resulting fusion algebra is again infinitely generated and non-commutative, and in a similar manner to the two dimensions, we can find the closed subalgebra by defining the dressed duality defect $\hatDuality{\bm{t}_1,\bm{t}_2}$ as follows; 
\begin{align}
    \hatDuality{\bm{t}_1,\bm{t}_2}(M_3)\equiv \etaSp\, (\bm{t}_1 )\times\CD_{\bm{t}_2} (M_{3})\times \CS\ ,
\end{align}
where $\bm{t}_{1}$ and $\bm{t}_{2}$ denote the homology cycles belonging to $H_{2}(M_3 , \BZ_{2N'})$, and 
\begin{align}
    \etaSp\, (\bm{t}_1)&=\exp\left[-\frac{ p^{K}\, \SlS_{IK}}{2N'}\int_{\bm{t}_1} \CG_{IJ}\ast dA^J\right]\ , \\
    \Duality_{\bm{t}_2}(M_3)&\equiv \exp\left[-\frac{\i}{2\pi}\int_{M_3}\,K_{IJ}\, A^{I}_{\text{L}}\, d A_{\text{R}}^{J}
    -\frac{\i \, p^{J}K_{IJ}}{2N'}\, \int_{\bm{t}_2} dA^I_{\text{L}}\right] \ .
\end{align}
We also have the orientation-reversed duality defect $\BarhatDuality{\bm{t}_1,\bm{t}_2}(M)\equiv\widehat{\CD}_{\bm{t}_1,\bm{t}_2}(\overline{M})$ as follows;
\begin{align}
    \BarhatDuality{\bm{t}_1,\bm{t}_2}=\hatDuality{-\bm{t}_2,\bm{t}_1}\times\CU=\CU\times\hatDuality{\bm{t}_2,-\bm{t}_1}\, .    
\end{align}
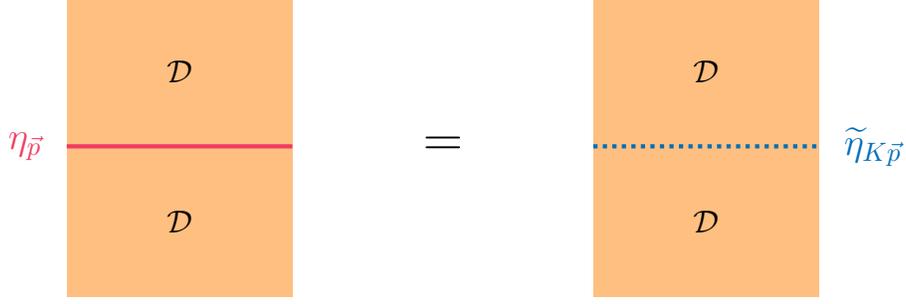
\begin{figure}[t]\label{fig:D times e in 4d}
   \centering
   \begin{tikzpicture}
        \fill[orange!50](-5, 0)--(-2, 0)--(-2, 4)--(-5,4)--cycle;
         \draw[color2, ultra thick](-5, 2)--(-2, 2);
          \fill[orange!50](2, 0)--(5, 0)--(5, 4)--(2 , 4)--cycle;
        \draw[color3, ultra thick,dotted](2, 2)--(5, 2);
         \node at (-3.5, 3) {{\large $\CD$}};
         \node at (-3.5, 1) {{\large $\CD$}};
         \node at (3.5, 3) {{\large $\CD$}};
         \node at (3.5, 1) {{\large $\CD$}};
         \node[left, color2] at (-5.2, 2) {{\Large $\eta_{\vec{p}}$}};
         \node[right, color3] at (5.2, 2) {{\Large $\widetilde{\eta}_{K \vec{p}}$}};
      \node at (0, 2) {{\huge $=$}};
   \end{tikzpicture}
   \caption{The result of the fusion rule $\CD \times \eta_{\vec{p}}$. Here, we imagine that the duality defect $\CD$ (orange plane) is sitting on this paper. In the left (right) diagram, the electric (magnetic) symmetry defect $\eta_{\vec{p}}$ ($\widetilde{\eta}_{K \vec{p}}\equiv \exp(\frac{\i\, p^{J}K_{IJ}}{2N'}\int dA^I)$) is living in the front (back) side of the duality defect. Both symmetry defects serve like the 1-morphisms, mapping from the below duality defect $\CD$ to the above one $\CD$.
   }
   \label{fig:higher category morphism}
 \end{figure}
Accordingly, we must also dress the condensation defect \cite{Kong:2014qka,Else:2017yqj,Gaiotto:2019xmp,Kong:2020cie,Kong:2020wmn,Johnson-Freyd:2020twl,Roumpedakis:2022aik} as follows;
\begin{align}\label{eq: projection operator}
        \begin{aligned}
        \widehat{\CC}_{\, \bm{t}_1,\bm{t}_2}(M_3)\equiv \exp\left[\frac{-\i p^{J} K_{IJ}}{2N'}\int_{\bm{t}_1}dA_{\text{L}}^{I}-\frac{  p^{K}\SlS_{JK} }{2N'}\int_{\bm{t}_2}\CG_{JM} \ast dA_{\text{L}}^{M}\right]\,\\
        \times\int \CD a \exp\left[-\frac{\i}{2\pi}\int_{M_3} K_{IJ}(A_{\Left}^I-A_{\Right}^I) d a^J\,\right]\, .   
        \end{aligned}
\end{align}
Then, we can obtain a {\it non-commutative} fusion subalgebra concerning $\hatDuality{\bm{t}_1,\bm{t}_2}$, $\widehat{\CC}_{\bm{t}_1,\bm{t}_2}$ and the $(\BZ^{[1]}_{2N'})_{\text{diag}}$ symmetry generator $\etap$ as follows;\footnote{We have checked that the obtained fusion subalgebra satisfies the associativity condition.}
\begin{shaded}
    \noindent \textbf{Non-commutative fusion subalgebra in pure U(1)$\times$U(1) gauge theory.} 
    \begin{align}\label{eq: fusion category four-dimension}
        \begin{aligned}
&\hatDuality{\bm{t}_1,\bm{t}_2}\times\BarhatDuality{\bm{t}_3,\bm{t}_4}=\widehat{\CC}_{-\bm{t}_2+\bm{t}_4 ,\bm{t}_1-\bm{t}_3}\ ,\\      &\BarhatDuality{\bm{t}_1,\bm{t}_2}\times\hatDuality{\bm{t}_3,\bm{t}_4}=\widehat{\CC}_{-\bm{t}_1+\bm{t}_3 ,-\bm{t}_2+\bm{t}_4}\ ,\\
       &\BarhatDuality{\bm{t}_1,\bm{t}_2}=\hatDuality{-\bm{t}_2,\bm{t}_1}\times\CU=\CU\times\hatDuality{\bm{t}_2,-\bm{t}_1}\ ,\\
      &\hatDuality{\bm{t}_1,\bm{t}_2}\times\eta_{\vec{p}}=\eta_{\vec{p}}\times\hatDuality{\bm{t}_1,\bm{t}_2}=\hatDuality{\bm{t}_1,\bm{t}_2}\ ,\\
      &\widehat{\CD}_{\bm{t}_1,\bm{t}_2}\times \widehat{\CC}_{\bm{t}_3,\bm{t}_4}=\CZ\, \hatDuality{\bm{t}_1+\bm{t}_3,\bm{t}_2+\bm{t}_4} \ ,\\ 
      &\widehat{\CC}_{\bm{t}_1,\bm{t}_2}\times \widehat{\CD}_{\bm{t}_3,\bm{t}_4}=\CZ\, \hatDuality{\bm{t}_2+\bm{t}_3,-\bm{t}_1+\bm{t}_4}\ , \\      &\widehat{\CC}_{\bm{t}_1,\bm{t}_2}\times\widehat{\CC}_{\bm{t}_3,\bm{t}_4}=\CZ\,\widehat{\CC}_{\bm{t}_1+\bm{t}_3,\bm{t}_2+\bm{t}_4}\ ,\\ 
        &\eta_{\vec{p}}^{2N'}=1\ ,
        \end{aligned}
    \end{align}
     where $\CZ$ is the decoupled topological field theory defined by;
    \begin{align}
        \CZ(M_3)\equiv\int\CD a\CD b\exp\left[-\frac{\i}{2\pi}\int_{M_3} K_{IJ}\, a^Idb^J\right]\ .
    \end{align}
    Here, $a$ and $b$ are dynamical U(1) gauge fields living in $M_3$, decoupled from the bulk theory. 
\end{shaded}
Finally, we give some comments on the fusion rule between the duality defect $\CD$ and the $(\BZ^{[1]}_{2N'})_{\text{diag}}$ symmetry generator $\eta_{\vec{p}}$. The fusion rule $\Duality\times\eta_{\vec{p}}$ can be evaluated as follows; 
\begin{align}\label{eq:D times eta in 4d} 
\begin{aligned}
    \Duality(M_{3})\times\eta_{\vec{p}}(\Sigma_{2})
    &=\exp\left(-\frac{\, p^{I}}{2N'}\int_{\Sigma_{2}} \CG_{IJ} \star dA^{J}_{\text{R}}-\frac{\i}{2\pi}\int_{M_3}\,K_{IJ}\, A^{I}_{\text{L}}\, d A_{\text{R}}^{J}\right) \\
    &=\exp\left(-\frac{\i \,  p^{J}K_{IJ}}{2N'}\int_{\Sigma_2}\,  dA^I_{\text{L}}-\frac{\i}{2\pi}\int_{M_3}\,K_{IJ}\, A^{I}_{\text{L}}\, d A_{\text{R}}^{J}\right)\, .
\end{aligned}
\end{align}
 From the right-hand side, the electric and magnetic one-form symmetry defects look like the 1-morphisms which map the duality defect $\CD$ to the same one in the context of the higher category \cite[section 2]{Bhardwaj:2022yxj}. 
Our ``1-morphism'' is, however, different from the standard 1-morphism in the higher category. This is because the different 1-morphisms, namely electric and magnetic one-form symmetry defects, appear on the left and right sides of the duality defect $\CD$. It may be an interesting open question to explore the mathematical structures of our ``1-morphism''.

\section{Conclusion and outlook}\label{sec: conclusion and discussion}
In this paper, we explored the non-invertible symmetries by using the half-space gauging associated with the diagonal sub-group $(\BZ_{2N'}^{[q]})_{\text{diag}}$. In the $c=2$ bosonic torus CFT, we showed that the diagonal gauging produces the non-invertible symmetry on the irrational CFT point, and derived the fusion algebra. In order for the algebra to close, we need to dress the duality defect with various global symmetry generators, and the resulting fusion algebra is non-commutative. Also, we apply the half-space gauging to the pure U(1)$\times$U(1) gauge theory in four dimensions and discuss the fusion algebra in a very similar manner to the $c=2$ bosonic torus CFT.
The consequent fusion algebra in four dimensions is also non-commutative.

We conclude this paper by mentioning some future directions;
\begin{itemize}
    \item In this paper, we mainly focused on the diagonal gauging, yet we can consider other gaugings. For instance, we can consider the gauging condition $p^1 n_1 +p^2 n_2 =0 \mod N$, which is more general compared with the diagonal gauging. Also, shift and winding symmetries in two dimensions (correspondingly, electric and magnetic one-form symmetries in four dimensions) have a mixed ’t Hooft anomaly. Hence, they cannot be gauged simultaneously and we cannot naively implement the half-space gauging associated to them. Even in that case, however, if we choose the nice discrete subgroup, we are free from any 't Hooft anomalies and can proceed with the half-space gauging\footnote{We thank Kantaro Ohmori for pointing out this possibility.} \cite{Cordova:2023ent}. The half-space gauging via these other gaugings may result in new non-invertible symmetries which are not captured in this paper.
    \item In more general, we can add the topological terms by turning on the B-field and theta angle in two and four dimensions, respectively. In this paper, we only consider the case where the charge lattice after gauging is a rectangular type. However, if we include such topological terms in the actions, the charge lattice is not limited to the rectangular one. This is because the B-field or theta angle makes the axis of the charge lattice tilted. It is interesting to investigate this generalization.  
\end{itemize}
Addressing these future directions would help with completing the landscape of non-invertible symmetries in the $c=2$ bosonic torus CFT and the pure U$(1)\times$U$(1)$ gauge theory, and we leave them to intriguing avenues for future works.

\vspace{-2mm}
\acknowledgments
\vspace{-3mm}
We are particularly grateful to Justin\,Kaidi, Kantaro\,Ohmori and Satoshi\,Yamaguchi for many enlightening comments and discussions. We are also grateful to Takamasa\,Ando, Yuma\,Furuta, Yui\,Hayashi, Hiroki\,Imai, Hayato\,Kanno, Kohki\,Kawabata, Ryutaro\,Matsudo, Tatsuma\,Nishioka for valuable discussions. Discussions during the YITP workshop on “Strings and Fields 2023” were useful to complete this work.
This work of Y.\,N. was supported by JST SPRING, Grant Number JPMJSP2138.
The work of S.\,S. was supported by JSPS fellowship for young students, Grant Number 23KJ1533.

\appendix
\section{Derivations of the selected fusion algebras in section \ref{sec:Fusion rules}}\label{sec: derivation of fusion algebra}
In this appendix, we provide concrete calculations, mainly focusing on the fusion algebras, which are omitted in section \ref{sec:Fusion rules}. Our methodology to derive the fusion rules closely follows the work \cite[section 6]{Choi:2022zal}. 
\begin{itemize}
    \item $\eta_{\vec{p}}\times \CD=\CD$\,  \eqref{eq: eta times D 2}  \vspace{2mm} \\
In order to derive the fusion rule $\eta_{\vec{p}}\times\Duality$, it is sufficient to consider the left bulk and the defect actions, which are given by
\begin{align}\label{eq: fusion rule eta times D}
    \frac{1}{4\pi}\int_{x<0}G^{*}_{IJ}\, d\phi^{I}_{\text{L}}\wedge\star \, d\phi^{J}_{\text{L}}+ \frac{\i}{2\pi}\int_{x=0} K_{IJ}\, \phi_{\text{L}}^{I}\, d \phi_{\text{R}}^{J}+\frac{p^{I}}{2N'} \int_{x=0}\, G^{*}_{IJ}\star d\phi_{\text{L}}^J\, .
\end{align}
By changing the path integral variable $\phi_{\text{L}}^{J}$;
\begin{align}\label{eq:phi_L shift}
\phi^{J}_{\text{L}}=\phi'^{J}_{\text{L}}-\frac{2\pi p^{J}}{2N'}\theta(-x)\, ,
\end{align}
where $\theta(x)$ is a step function defined as $\theta(x)=0$ in $x\leq 0$ and $\theta(x)=1$ in $x>0$, 
the above combined action \eqref{eq: fusion rule eta times D} becomes; 
\begin{align}\label{eq:defect action of e times D}
 \frac{1}{4\pi}\int_{x<0}G^{*}_{IJ}\, d\phi'^{I}_{\text{L}}\wedge\star \, d\phi'^{J}_{\text{L}}+
            \frac{\i}{2\pi}\int_{x=0}\,K_{IJ}\, \phi'^{I}_{\text{L}}\, d \phi_{\text{R}}^{J}
             -\frac{\i p^{I}K_{IJ}}{2N'}\int_{x=0}\,  d\phi^{J}_{\text{R}}\,.
\end{align}
 Note that $p^IK_{IJ}=0 \mod{2N'}$, therefore the last term can be dropped from the action.
Then, we get the following fusion rule;
\begin{align}\label{eq:D0 times eta}
     \eta_{\vec{p}}\times\Duality=\Duality\, .
\end{align}

    \item $\CD \times \eta_{\vec{p}}$\,  \eqref{eq: D times eta} \vspace{2mm} \\ 
As stressed in the main text, the fusing the $(\BZ_{2N'}^{[0]})_{\text{diag}}$ shift symmetry defect $\eta_{\vec{p}}$ to the duality defect $\Duality$ from the right shows a different behavior from the case of $\eta_{\vec{p}}\times\Duality$. To see this, we only have to consider the right bulk and the defect actions;
\begin{align}\label{eq: D times eta action}
 \frac{1}{4\pi}\int_{0<x}G^{*}_{IJ}\, d\phi^{I}_{\text{R}}\wedge\star \, d\phi^{J}_{\text{R}}+
    \frac{\i}{2\pi}\int_{x=0}\ K_{IJ}\, \phi_{\text{L}}^{I}\, d \phi_{\text{R}}^{J}+\frac{ p^{I}}{2N'}\int_{x=0}G_{IJ}\star \, d\phi_{\text{R}}^J\, .
\end{align}
We perform the following field redefinition:
\begin{align}\label{eq:phi_R shift}
        \phi^{J}_{\text{R}}=\phi'^{J}_{\text{R}}+\frac{2\pi p^{J}}{2N'}\theta(x)\, .
\end{align}
Then, the composite action \eqref{eq: D times eta action} can be calculated as follows;
\begin{align}\label{eq:defect action of D times e}
        \frac{1}{4\pi}\int_{0<x}G^{*}_{IJ}\, d\phi'^{I}_{\text{R}}\wedge\star \, d\phi'^{J}_{\text{R}}+\frac{\i}{2\pi}\int_{x=0}\,K_{IJ}\, \phi^{I}_{\text{L}}\, d \phi_{\text{R}}'^{J}
        -\frac{\i p^{J}K_{IJ}}{2N'}\int_{x=0}\,  d\phi^{I}_{\text{L}}\,,    
\end{align}
which shows that the non-trivial $\BZ_{2N'}$ winding symmetry generator appears on the left side of the duality defect $\CD$ due to $p^{J}K_{IJ}\not= 0 \mod 2N'$.
\end{itemize}
The following fusion rules are related to our main result \eqref{eq: fusion category}.
\begin{itemize}
    \item $\widehat{\CD}_{s_1,s_2} \times \widehat{\CD}_{s_3,s_4}=\widehat{\CC}_{s_2+s_3,s_1+s_4}$\vspace{2mm} \\ 
    By using the associative property of symmetry generators, the fusion rule $\widehat{\CD}_{s_1,s_2} \times \widehat{\CD}_{s_3,s_4}$ can be reduced to as follows;
    \begin{align}\label{eq: fusion rule hatd hatd}
    \begin{aligned}
                \widehat{\CD}_{s_1,s_2} \times \widehat{\CD}_{s_3,s_4}
        &=(\etaSp)^{s_1}\times(\CD_{s_2}\times \CS)\times (\etaSp)^{s_3} \times (\CD\times \CS)\times(\etaSp)^{s_4}\\
        &=(\etaSp)^{s_1}\times\CD_{s_2+s_3}\times (\CS\times  \CD\times \CS)\times(\etaSp)^{s_4}\\
        &=(\etaSp)^{s_1}\times\CD_{s_2+s_3}\times \overline{\CD}\times\etaSpb^{s_4}\ . 
    \end{aligned}
    \end{align}
     In the first line, we used the definition of the dressed duality defect $\widehat{\CD}_{s_1 , s_2}$ and \eqref{eq:relation of etaSp}. We also made use of \eqref{eq:relation of etaSp} and \eqref{eq:barduality relation} in the second and final lines, respectively.
    As easily can be checked, the fused defect action reads
    \begin{align}\label{eq: DDbar}
      \CD_{s_2+s_3}\times \overline{\CD}\ : \ \frac{\i (s_2+s_3)p^{J}K_{IJ}}{2N'}\int_{x=0}d\phi_{\text{L}}^{I}+\frac{\i}{2\pi}\int_{x=0}\,K_{IJ}\, (\phi_{\text{L}}^{I}-\phi_{\text{R}}^{I})\, d \varphi^{J}\, .
    \end{align}
     The first term is nothing but the $\BZ_{2N'}$ winding symmetry generator, which originates from the dressed duality defects $\widehat{\CD}_{s_2}$ and $\widehat{\CD}_{s_3}$. The second term is the projection operator associated with the $\BZ_{2N'}$ shift symmetry.
     To see this, we decompose the charge matrix $K$ into the Smith normal form;
    \begin{align}
    \begin{pmatrix}
    N' & -1 \\
    N' & 1 
    \end{pmatrix}
    =
    \begin{pmatrix}
    1 & -1 \\
    0 & 1 
    \end{pmatrix}
    \begin{pmatrix}
    2N' & 0 \\
    0 & 1 
    \end{pmatrix}
    \begin{pmatrix}
    1 & 0 \\
    -N' & 1 
    \end{pmatrix}
    .
    \end{align}
    This shows that the second term can be split into the $\BZ_{2N'}$ BF theory and the trivial one, and we can write it as a sum of $\BZ_{2N'}$ generators (See \cite[Appendix E]{Niro:2022ctq} for the derivation).
    Also, as we attached the $\BZ_{2N'}$ element to the duality defect, we define the dressed projection operator $\widehat{\CC}_{s_1 , s_2}$ $(s_1 , s_2 =0, 1, \cdots 2N'-1)$ as follows;
     \begin{align}\label{eq: projection operator}
     \begin{aligned}
        \widehat{\CC}_{s_1,s_2}(\Sigma)\equiv &\exp\left[-\frac{\i s_1 p^{J} K_{IJ}}{2N'}\int_{\Sigma} d\phi_{\text{L}}^{I}\right]\,\\
        &\times\int \CD\varphi \exp\left[-\frac{\i}{2\pi}\int_{\Sigma} K_{IJ}\left(\phi_{\Left}^I-\phi_{\Right}^I-\frac{2\pi s_{2}}{2N'}\SlS^{IK}p_{K}\right)d\varphi^J\,\right]\,.
     \end{aligned}
    \end{align}
    By using this dressed projection operator, the fusion rule $\CD_{s_2+s_3}\times \overline{\CD}$  can be written as 
    \begin{align}
        \CD_{s_2+s_3}\times \overline{\CD}=\widehat{\CC}_{s_2+s_3,0}\,.
    \end{align}
    Then, we can evaluate the fusion rule $\widehat{\CD}_{s_1,s_2} \times \widehat{\CD}_{s_3,s_4}$ as follows;
    \begin{align}
        \widehat{\CD}_{s_1,s_2} \times \widehat{\CD}_{s_3,s_4}=\etaSpb^{s_1}\times\widehat{\CC}_{s_2+s_3,0}\times\etaSpb^{s_4}=\widehat{\CC}_{s_2+s_3,s_1+s_4}\ .
    \end{align}
    In the last equality, we used the following fusion rule: 
    \begin{align}
\widehat{\CC}_{s_1,s_2}\times\etaSpb^{s_3}=\etaSpb^{s_3}\times\widehat{\CC}_{s_1,s_2}=\widehat{\CC}_{s_1,s_2+s_3}\ , 
    \end{align} 
    which can be understood as follows. After redefining the integral variables as follows:
    \begin{align}
\phi^{J}_{\text{R}}=\phi'^{J}_{\text{R}}+\frac{2\pi }{2N'}\SlS^{JK}p^{K}\theta(x)\,,
    \end{align}
     it turns out that the fusing $\etaSpb^{s_3}$ to $\widehat{\CC}_{s_1,s_2}$ from the right bulk results in the shift of $\phi_{\text{R}}^{I}$ on $\Sigma$ by $\frac{2\pi s_{3}}{2N'}\SlS^{IK}p_{K}$. This is equivalent to the replacement the label $s_{2}$ in the dressed projection operator $\widehat{\CC}_{s_1,s_2}$ as $s_{2}\rightarrow s_{2}+s_{3}$, hence the fusion rule $\widehat{\CC}_{s_1,s_2}\times\etaSpb^{s_3}=\widehat{\CC}_{s_1,s_2+s_3}$ can be deduced. In a very similar manner, we can also derive the fusion rule $\etaSpb^{s_3}\times\widehat{\CC}_{s_1,s_2}=\widehat{\CC}_{s_1,s_2+s_3}$\footnote{Here, we should note that shift and winding symmetry generators commute with each other. }.

     \item $\eta_{\vec{p}}\times \widehat{\CD}_{s_1,s_2}=\widehat{\CD}_{s_1,s_2}$ and $\widehat{\CD}_{s_1,s_2} \times \eta_{\vec{p}}=\widehat{\CD}_{s_1,s_2}$\vspace{2mm}\\ 
    Here, we derive the fusion rules $\eta_{\vec{p}}\times \widehat{\CD}_{s_1,s_2}$ and $\widehat{\CD}_{s_1,s_2} \times \eta_{\vec{p}}$ by using the associativity. The fusion $\eta_{\vec{p}}\times \widehat{\CD}_{s_1,s_2}$ can be easily evaluated as follows;
     \begin{align}
     \begin{aligned}
        \eta_{\vec{p}}\times \widehat{\CD}_{s_1,s_2}
  &=\eta_{\vec{p}}\times(\etaSp)^{s_1}\times\CD_{s_2}\times\CS \\
  &= (\etaSp)^{s_1}\times\eta_{\vec{p}}\times\CD_{s_2}\times\CS\\
  &= (\etaSp)^{s_1}\times\CD_{s_2}\times\CS \\
        &=\widehat{\CD}_{s_1,s_2}\, , 
     \end{aligned}
    \end{align}
   where in the third line, we used the fusion rule \eqref{eq:e times Ds}.
    Likewise, we can derive the fusion rule $\widehat{\CD}_{s_1,s_2} \times \eta_{\vec{p}}$ as follows; 
        \begin{align}
        \begin{aligned}
            \widehat{\CD}_{s_1,s_2} \times \eta_{\vec{p}}
            &=(\etaSp)^{s_1}\times\Duality\times\eta_{\vec{p}}^{s_2}\times\CS\times\eta_{\vec{p}}\\
            &=(\etaSp)^{s_1}\times\Duality\times\eta_{\vec{p}}^{s_2}\times\eta_{\, \SlS\vec{p}}\times\CS\\
            &=(\etaSp)^{s_1}\times\Duality\times\eta_{\, \SlS\vec{p}}\times\eta_{\vec{p}}^{s_2}\times\CS\\
            &=(\etaSp)^{s_1}\times\Duality\times\eta_{\vec{p}}^{s_2}\times\CS \\
            &= \widehat{\CD}_{s_1,s_2}\ .
        \end{aligned}
        \end{align}
         In the fourth line, we used the following formula;
        \begin{align}
        \Duality\times \eta_{\, \SlS\vec{p}}=\Duality\,.
    \end{align}
    which can be derived by a similar way to \eqref{eq: eta times D} and \eqref{eq: D times eta}.
     \item $\hatDuality{s_1,s_2}\times\widehat{\CC}_{s_3,s_4}$ \vspace{2mm}\\ 
    We can rewrite $\hatDuality{s_1,s_2}\times\widehat{\CC}_{s_3,s_4}$ as follows;
    \begin{align}
    \hatDuality{s_1,s_2}\times\widehat{\CC}_{s_3,s_4}
    &=\etaSpb^{s_1}\times\Duality\times\CS\times\etaSpb^{s_2}\times\widehat{\CC}_{s_3,s_4}\\
    &=\etaSpb^{s_1}\times\CS\times\overline{\Duality}\times\widehat{\CC}_{s_3,0}\times\etaSpb^{s_2+s_4}\,.
\end{align}
In the last line, we used the following relation;
\begin{align}\label{eq:etaSp times CC}
    \etaSpb^{s_2}\times\widehat{\CC}_{s_3,s_4}=\widehat{\CC}_{s_3,s_2+s_4}=\widehat{\CC}_{s_3,0}\times \etaSpb^{s_2+s_4}\,.
\end{align}
Therefore, in order to calculate the fusion rule $\hatDuality{s_1,s_2}\times\widehat{\CC}_{s_3,s_4}$, we firstly need to derive the fusion rule $\overline{\Duality}\times\widehat{\CC}_{s_3,0}\,$.
The defect action is given by
\begin{align}
    \overline{\Duality}\times\widehat{\CC}_{s_3,0}\,:\,\frac{\i s_3p^{J}}{2N'}\int_{x=0} K_{IJ}d\phi_{\text{M}}^I+\frac{\i}{2\pi}\int_{x=0} K_{IJ}\,(\phi_{\text{M}}^I-\phi_{\text{R}}^I)d\varphi^J+\frac{\i}{2\pi}\int_{x=0} K_{JI}\,\phi_{\text{L}}^Id\phi_{\text{M}}^J\,.
\end{align}
By changing the path integral variables as follows: 
\begin{align}
    \phi'^{I}_{\text{M}}=\phi_{\text{M}}^I-\phi_{\text{R}}^I\qquad , \qquad \varphi'^{I}=\varphi^I-\phi_{\text{L}}^I\ , 
\end{align}
and the defect action becomes;
\begin{align}
    \frac{\i}{2\pi}\int_{x=0} K_{IJ}\, \phi'^{I}_{\text{M}}d\varphi'^{J}+\frac{\i s_3p^{J}}{2N'}\int_{x=0} K_{IJ}\,d\phi'^{I}_{\text{M}}+\frac{\i}{2\pi}\int_{x=0} K_{JI}\,\phi_{\text{L}} ^{I}d\phi_{\text{R}}^{J}+\frac{\i s_3 p^{J}}{2N'}\int_{x=0}\, K_{IJ}d\phi_{\text{R}}^I\,.
\end{align}
The first two terms represent the decoupled TQFT $\CZ$, and this can be written as the sum of $\BZ_{2N'}$ symmetry generators. Hence, we have $\CZ=2N'$. Also, the last two terms are the defect actions of $\etap^{s_3}\times\overline{\Duality}\,$ since the winding symmetry generator is changed to the shift one across the duality defect. Combining the above results, the fusion rule $\hatDuality{s_1,s_2}\times\widehat{\CC}_{s_3,s_4}$ can be derived as follows;
\begin{align}\label{eq:hatD times CC}
\begin{aligned}
        \hatDuality{s_1,s_2}\times\widehat{\CC}_{s_3,s_4} &=2N'\etaSpb^{s_1}\times\CS\times\etap^{s_3}\times\overline{\Duality}\times\etaSpb^{s_2+s_4}\\
&=2N'\etaSpb^{s_1+s_3}\times\CD\times\CS\times\etaSpb^{s_2+s_4}\\
&=2N'\etaSpb^{s_1+s_3}\times\CD\times\etap^{s_2+s_4}\times\CS\\
    &=2N'\hatDuality{s_1+s_3,s_2+s_4}\,.
\end{aligned}
\end{align}
 We can also calculate the fusion rule $\widehat{\CC}_{s_1,s_2}\times\hatDuality{s_3,s_4}$ in a similar manner to the above derivation.
\end{itemize}

\bibliographystyle{JHEP}
\bibliography{Non-inv}

\end{document}